\documentstyle[amssymb,aps,multicol,epsf,pre]{revtex}

\def\beginwide{
        \end{multicols} \vspace*{-0.5cm} \noindent
        \rule{3.5in}{.1mm}\rule{.1mm}{5mm} \widetext \medskip }
\def\endwide{
        \hspace*{3.5in}~\rule[-5mm]{.1mm}{5mm}\rule{3.5in}{.1mm}
        \begin{multicols}{2}\narrowtext \vspace*{-1.0cm} \noindent }

\begin{document}

\title{Driven Dynamics of Periodic Elastic Media in Disorder}

\author{Stefan Scheidl$^{1,2}$ and Valerii M. Vinokur$^2$}

\address{$^1$Institut f\"ur Theoretische Physik, Universit\"at zu
  K\"oln, Z\"ulpicher Str. 77, D-50937 K\"oln, Germany\\ $^2$Materials
  Science Division, Argonne National Laboratory, Argonne, IL 60439}

\date{\today}

\maketitle

\begin{abstract}
  We analyze the large-scale dynamics of vortex lattices and charge
  density waves driven in a disordered potential. Using a perturbative
  coarse-graining procedure we present an explicit derivation of
  non-equilibrium terms in the renormalized equation of motion, in
  particular Kardar-Parisi-Zhang non-linearities and dynamic strain
  terms.  We demonstrate the absence of glassy features like diverging
  linear friction coefficients and transverse critical currents in the
  drifting state. We discuss the structure of the dynamical phase
  diagram containing different elastic phases very small and very
  large drive and plastic phases at intermediate velocity.
\end{abstract}

\pacs{PACS numbers: 05.70.Ln, 71.45.Lr, 74.25.Dw, 74.40.+k}

\begin{multicols}{2} \narrowtext
\section{Introduction}

Periodic structures driven through a random environment have become a
paradigm for the statistical mechanics of non-equilibrium processes.
The beginning of the study of this phenomenon in the context of charge
density wave (CDW) dynamics was marked by the development of several
pioneering and elegant concepts (see \cite{cdw} for a review), in
particular the description of the depinning transition in terms of
critical phenomena \cite{fisher}. Yet it was hard to foresee the
subsequent growth of what seemed to be a simple although a subtle
subject into a fascinating multidisciplinary branch of statistical
physics.  The resurgence of interest was related to the discovery of
high temperature superconductors (HTS) where the motivation was driven
also by the technological quest for the description of transport
properties of HTS.

The understanding of the remarkable effects displayed by driven vortex
lattice involved a diversity of concepts drawn from various branches
of contemporary physics ranging from polymer physics and spin glasses
to nonlinear stochastic equations and turbulence, as well as the
invention of new concepts of non-equilibrium physics of disordered
media (see \cite{Bla+94}). In recent years much theoretical effort has
been expended to advance our knowledge of the driven dynamics of
disordered media.  Yet in spite of impressive achievements there
remains a vast amount of fundamental open questions with the depth and
subtleties still to be revealed. In this work we develop a regular
approach to the description of periodic media driven through a
quenched random environment that will hopefully enable to put
subsequent research endeavors on a firm standard basis.

\bigskip

{\it Statics of disordered elastic periodic systems}

\bigskip

The subtle dynamic properties of dirty media are governed by the
interplay among thermal fluctuations, driving force, and quenched
disorder.  To gain better insight into the dynamics we first discuss
briefly the {\it statics} of weakly disordered elastic periodic
systems. These include CDW, vortex lattices (VL), vortex arrays in
Josephson junctions, domain walls, dislocations in solids, Wigner
crystals, and many others.  The common feature of the above systems is
that, although the weakness of pinning suggests the purely elastic
Hamiltonian as a starting point, the disorder-distorted system
possesses a huge number of meta-stable states and the ground state is
infinitely degenerate. This dooms to eventual failure direct attacks
on the asymptotic large-scale behavior based on a straightforward
perturbation theory with respect to disorder.

The first and decisive step for an approach by statistical physics to
such systems was made in the remarkable work of Larkin \cite{aiL70}.
The pioneering ideas of this work were later cast into the {\it
collective pinning theory} \cite{lo73,fl78,lo79} and basically
determined the further development of the field.  It was recognized in
\cite{aiL70} that pinning can be treated perturbatively in the domain
of the distorted lattice belonging to a single meta-stable state
generated by disorder.  Such a coherently pinned domain is called the
correlated volume and the pinning energy stored in such a domain
determines the crucial characteristic of the pinned system: the
critical depinning force.

The key quantity characterizing the system is the degree of distortion
of the elastic system by disorder, the {\it roughness} $w({\bf r})=
\overline{\langle [{\bf u}({\bf r})-{\bf u}({\bf 0})]^2\rangle }$,
where ${\bf u}({\bf r})$ is the displacement of a vortex from its
undistorted position ${\bf r}$. Within the domain $w(r)\leq\xi^2$,
where $\xi$ is the characteristic spatial scale of variations of the
random potential, the pinning force ${\bf f}$ has only a negligible
dependence on the displacements ${\bf u}$.  This implies that this
domain is pinned coherently and lies in the single valley of the
effective potential landscape of the system. The roughness within the
correlation volume grows as $w(r) \propto r^{2 \zeta}$ with the
so-called wandering or roughness exponent $\zeta$ that takes the value
$\zeta = \frac{4-d}2$ in the Larkin regime ($d$ is the dimensionality
of the lattice). Since different Larkin domains are pinned
independently one could conclude that pinning, however weak, destroys
the long-range order of the lattice for $d<4$.

As soon as the relative displacement of vortices exceeds the disorder
correlation length, the spatial variation of the pinning forces
becomes important. The vortices start to feel that they are in a
random {\em potential} having many meta-stable states rather than in a
random force field. Therefore the perturbative result would
overestimate the actual roughness of the lattice. On intermediate
scales where $\xi^2 \leq w(r) \leq a^2$ and $a$ is the lattice
constant the system adjusts itself to the multi-valley potential
relief. This regime is referred to as the random manifold regime
\cite{feig} and the roughness exponent becomes smaller than in the
Larkin regime. On these scales the periodicity of the medium is not
yet of significance. The region of the largest scales, where lattice
displacements exceed the lattice constant, $a^2\leq w(r)$, and the
periodic nature comes into play, was first investigated by Nattermann
\cite{N90} who found, by the renormalization group approach, that at
large distances pinning of the VL is equivalent to pinning of CDW and
that displacements grow only logarithmically, $w(r)\propto \ln
(r)$. The above results were confirmed later by variational replica
approaches \cite{BMY,korsh93,GD94,GD95}.

Structures with logarithmic roughness are well known in the physics of
surfaces and 2D crystals.  The logarithmic roughness implies that the
system retains its periodic character and Bragg peaks in the structure
factor $S({\bf q})$, the singularities however have an algebraic
character, $S({\bf q})\propto |{\bf q}-{\bf Q}|^{-\nu}$, (${\bf Q}$ is
a reciprocal lattice vector) rather than the $\delta$-function-like
character as in ordinary crystals or Lorentzian character in liquids
\cite{merm,janc,ber}.  This algebraic behavior is a characteristic
feature of {\it quasi-long-range crystalline order}.

The roughness of the lattice structure implies a ruggedness of the
potential landscape of the system and the existence of infinitely high
barriers separating the different meta-stable states, which is the
characteristic feature of glassy systems (see \cite{Bla+94}).  This
was realized in a seminal work by M.P.A. Fisher \cite{vg} who
identified the VL distorted by disorder as a glassy structure and
called it the {\it vortex glass}.  It is important to stress that the
derivation of the above features was based on the {\it elastic} nature
of disordered lattices.

The stability of the elastic vortex glass with respect to the
formation of topological defects (dislocations) was questioned by
Fisher, Fisher and Huse \cite{ffh} who stated that dislocations are to
be generated at the scales where the roughness becomes of order the
vortex spacing and that therefore the elastic description of the
vortex glass fails. In spite of the fact that the correctness of the
arguments of \cite{ffh} was questioned in its turn (the energy of the
dislocation formation was underestimated and the logarithmic
smoothness of the lattice on large scales was overlooked), the image
of the vortex glass as a dislocation saturated medium became
widespread.  Arguments demonstrating the self-consistency of the
elastic vortex glass approach (as long as the disorder is weak enough)
were presented in \cite{GD95,ep-stat,f97}. Thus the existence of a
weak disorder-induced elastic vortex glass free of topological defects
can be considered as well established. Recalling that a
logarithmically rough medium shows algebraic Bragg peaks, Giamarchi
and Le Doussal \cite{GD95} proposed to call the vortex lattice
deformed by disorder a ``Bragg glass.''  This name gained popularity
among the specialists and replaced the somewhat compromised ``vortex
glass'' in their technical jargon. The latter terms is now reserved
for the topologically disordered vortex solid phase.

\bigskip

{\it Dynamics of disordered elastic periodic systems}

\bigskip

The main feature of the driven dynamics in a random environment is the
existence of the pinning threshold: at zero temperature the system
remains pinned if the drive does not exceed a certain threshold value
$f_c$, the critical pinning force, and slides if $f>f_c$.  At finite
temperatures the sharp transition is rounded by thermal fluctuations
and is not very well defined. Nevertheless, one can keep the notion of
a temperature dependent critical force $f_c(T)$ as the force
separating the pinning dominated regime with slow thermally activated
dynamics or creep at low force $f \ll f_c$ from the fast sliding one
at $f \gg f_c$.

Motion of the system in the creep regime occurs via thermally
activated jumps over energy barriers separating different meta-stable
states. The size of the typical energy barrier as a function of the
driving force can be related by scaling arguments to structural
features of the system, in particular its roughness. This was done in
\cite{iv87} for driven elastic manifolds and it was found that the
barriers controlling the motion {\it diverge} algebraically at small
driving forces as $U(f) \propto 1/f^{\mu}$, where $\mu= (2\zeta-1) /
(2-\zeta)$.  The approach of \cite{iv87} was extended from continuous
media to the creep of vortex lattices in \cite{feig,N90}. The
divergence of the activation barriers implies a non-linear response of
the system to small forces and leads to the identification of the
low-temperature vortex state as a glassy phase, since such a
non-linear response is a hallmark of the glassy system.

The understanding of the critical behavior at the depinning threshold
has seen remarkable progress\cite{depin,mid,TS} since the first work
by D.S. Fisher \cite{fisher}.

The high-velocity sliding of the periodic systems was long considered
as the most ``easy-to-understand'' regime.  The outburst of the
interest in the flow regime at large driving forces well above the
depinning threshold was triggered by the prediction \cite{KV94} of
dynamic phase transitions between plastic sliding in the nearest
vicinity of depinning and coherent motion of the crystalline structure
at high drives.  Already early experiments \cite{Tho+73} have shown
that a moving vortex lattice has a higher degree of crystalline order
than a pinned vortex lattice. These studies have been refined recently
\cite{Bhatt,Yaron} to identify the different dynamical regimes.

The non-equilibrium phase transition predicted in Ref. \cite{KV94} is
expected to occur in systems with sufficiently strong disorder, where
depinning is accompanied by the massive production of topological
defects \cite{jensen}.  The structural order improves at large driving
forces because the system experiences disorder forces that are
temporally fluctuating in the moving frame. In this sense the effect
of disorder resembles the thermal noise of a heat bath.  However, this
comparison does not carry too far, since the quenched nature of the
disorder still implies infinite-ranged spatio-temporal correlations of
the effective force in the moving frame.

Balents and Fisher \cite{BF95} used scaling arguments to extend the
concept of the non-equilibrium ``freezing'' transition of
Ref. \cite{KV94} to CDW and have shown that true long-range order is
restored at large velocities only in $d>3$. Thus the critical
dimension is reduced by one in comparison to the static case. For $d
\leq 3$ the CDW phase, which is the analog of the vortex displacement,
is still rough even even at the largest velocities, whereas a temporal
order resulting in narrow-band noise may still persist.

Giamarchi and Le Doussal \cite{GL96} addressed the question of spatial
large-scale order in the driven VL.  They argued that the transverse
periodicity of the system leads to glassy features of the driven
phase.  It was argued, in particular, that on the largest scales the
driven lattice retains a logarithmic roughness in the directions
transverse to the velocity for $2<d \leq 3 $.  In the direction
parallel to the velocity the roughness is expected to be even
algebraic as for driven CDW \cite{BMR97,GL97} in contrast to the
static case, where roughness is logarithmic. One of the manifestations
of the transverse glassiness suggested according to Ref. \cite{GL96}
would be the existence of diverging transverse barriers.  This brings
to mind the early observation of Schmid and Hauger \cite{SH73} who
have performed a lowest-order perturbative calculation for the pinning
force and noticed a discontinuity in the transverse I-V characteristic
in a sliding state.

Notice that in comparison to the equilibrium situation in the absence
of driving forces, the approach to the physics at large velocities is
even more intricate because of the {\em non-equilibrium} nature of the
driven state.  Although important predictions about this state have
already been formulated, a systematic approach is still lacking.

In this paper we develop such a systematic approach to the driven
dynamics of dirty periodic media on the basis of the
Martin-Siggia-Rose (MSR) formalism.  The MSR formalism provides a
powerful tool to access the largest scales and to treat the immediate
vicinity of the depinning transition.  Although the complete
description should include the derivation of renormalization group
equations, we will show that a number of important conclusions
concerning the properties of the driven state can be successfully
achieved even within the framework of the dynamic perturbative
approach. For large velocities $v$ the small parameter for the
expansion is $\Delta_0/(\eta^2 v^2 \xi^{d+2})$ where $\Delta_0$ is the
$d$-dimensional spatial integral of the potential correlator of a
width $\xi$, and $\eta$ is the friction coefficient.

\bigskip

{\it Summary of results}

\bigskip

Using a coarse-graining procedure for the dynamics of periodic media,
we find that their large-scale behavior is governed by an effective
equation of motion 
\begin{eqnarray}
  \label{eqmo^*}
  \eta_{\alpha \beta}^* \dot u_\beta &=& 
  D_{\alpha \beta}^* u_\beta + F_\alpha-  F_\alpha^{{\rm fr}*}
  + \frac 12 \lambda_{\alpha \beta \gamma ab}^*
  [\partial_a u_\beta][\partial_b u_\gamma]
  \nonumber \\ 
  && +\xi_\alpha^* +f_\alpha^* ({\bf r}+{\bf v}t + {\bf u}) ~.
\end{eqnarray}
Renormalized parameters carry an asterisk to distinguish them later on
from the unrenormalized (bare) ones.  Under this procedure the
parameters become anisotropic since the velocity identifies a
particular spatial direction.

All components of the friction coefficient $\eta^*$ are found to be
{\em finite}. Therefore glassy features, which in general appear as a
divergence of such coefficients, are {\rm absent}. The elastic
dispersion that reads $D^*_{\alpha \beta}({\bf q}) =-i \chi^*_{\alpha
\beta a} q_a + \kappa^*_{\alpha \beta ab} q_a q_b$ in Fourier space,
includes besides elastic constants $\kappa$ also stress terms $\chi$
after coarsening.

Due to pinning and dissipative effects on spatial scales smaller than
the coarse-grained cutoff scale (with momenta larger than
$\Lambda^<$), the velocity-dependence of the friction force $F^{{\rm
fr}*}$ becomes non-linear. 

The fourth term in Eq. (\ref{eqmo^*}) is a Kardar-Parisi-Zhang (KPZ)
non-linearity \cite{KPZ86} $\lambda^*$, which is absent in the bare
dynamics, is generated by disorder. It is an anisotopic generalization
of the term $\frac 12 \lambda ({\bbox \nabla} u)^2$ familiar from
surface growth and Burger's equation.

The pinning force $f^*$, which was simply the gradient of a random
potential in the bare case, acquires a more virulent random-force
character with a correlator
\begin{mathletters}
  \label{phi.xx.0}
  \begin{eqnarray}
	\phi^*_{\alpha \beta} &=& \int_{\bf R} 
  \overline{ f_\alpha^*({\bf R}) f^*_\beta({\bf 0}) }~, \\ 
  \phi^{*}_{xx} &\approx& \frac{\Delta_0^2}{ \xi^{4+d} \eta^2 v^2} ~,
  \\
    \phi^{*}_{yy} &\approx& \frac{a^2 \Delta_0^2}{\xi^{3+d}c \eta v}  
    ~ {\rm for} ~ d>3 ~, \\
    \phi^{*}_{yy} &\approx& \frac{a^2 \Delta_0^2}{\xi^{3+d}c \eta v}
    (a \Lambda^<)^{d-3}  ~ {\rm for} ~ d<3
  \end{eqnarray}
\end{mathletters}
for large velocities $v \gg c \xi/(\eta a^2)$ with a typical elastic
constant $c$. One sees that the variance of the components
perpendicular to the velocity {\em diverges} in $d \leq 3$ as the
coarse-graining cutoff $\Lambda^* \to 0$.

The effective thermal noise $\xi^*$ describes in general an effective
heat bath with a temperature that is increased due to  shaking
effects exerted by the pinning on the medium.  
\begin{mathletters}
  \label{vartheta.xx.intro}
  \begin{eqnarray}
    \vartheta_{\alpha \beta}^* &=&
    \int_{t \bf r} \langle \xi^*_\alpha({\bf r},t) 
    \xi^*_\beta({\bf 0},0) \rangle ~, \\
    \vartheta^{*}_{xx} &\approx& \vartheta + 
    \frac{\Delta_0}{ \xi^{2+d} \eta^2 v^2} 
    \vartheta ~,   \\
    \vartheta^{*}_{yy} &\approx& \vartheta + 
    \frac{a^2 \Delta_0}{\xi^{3+d}c \eta v}  
    \vartheta  ~ {\rm for} ~ d>3 ~, \\
    \vartheta^{*}_{yy} &\approx& \vartheta + 
    \frac{a^2 \Delta_0}{\xi^{3+d}c \eta v}
    (a \Lambda^<)^{d-3} \vartheta   ~ {\rm for} ~ d<3 ~.
  \end{eqnarray}
\end{mathletters}
In the non-equilibrium case this effective ``temperature'' is defined
as the integral over the correlator of $\xi^*$, that can be
distinguished from pinning forces by the temporal decay of its
correlations. The behavior of $\vartheta^*$ and $\phi^*$ is very
similar, they show the same type of divergence for $\Lambda^< \to 0$.
Since this divergence comes from small momenta, it is a measure for
the strong fluctuations of the medium on {\em large} scales only.

The disorder generated stress couplings $\chi$, the KPZ
non-linearities and the random forces are specific non-equilibrium
terms that are absent in the equation of motion before
coarse-graining.

From the effective equation of motion (\ref{eqmo^*}) the displacement
fluctuations are found to roughen the medium in dimensions $d \leq 3$
with high anisotropy for directions parallel and perpendicular to the
velocity (which we choose to be parallel to the $x$-axis). The
anisotropy manifests itself in two distinct features of the
displacement correlations: (i) The components of the displacement
field perpendicular to ${\bf v}$ exhibit stronger fluctuations than
the components parallel to ${\bf v}$. (ii) The relative displacement
increases faster in perpendicular directions than in the parallel
direction.

The former feature, obtained below within perturbation theory for the
elastic medium, can be understood qualitatively already from a
single-particle picture in analogy to the consideration that led to
the notion of the shaking temperature, describing the disorder-induced
increase of the effective system temperature \cite{KV94}. To this end
we consider a particle moving in a disorder potential $V$ with
Gaussian correlations $\overline{V({\bf R}) V({\bf 0})} \sim
{\Delta_0} \xi^{-d} e^{-R^2/(2 \xi^2)}$. A particle starting at ${\bf
R}(t=0)=0$ moves with an average velocity ${\bf v} \approx {\bf F} = F
{\hat {\bf x}}$ following an over-damped equation of motion $\eta \dot
{\bf R}={\bf F} - {\bbox \nabla} V({\bf R}) \approx \eta {\bf v} -
{\bbox \nabla} V({\bf v} t)$. The components of its displacement ${\bf
u}(t)={\bf R}(t)-{\bf v} t$ parallel and perpendicular to ${\bf v}$
have a variance that grows differently as a function of time. When the
effect of the pinning forces is integrated over time (i.e. along the
direction of motion of the particle), the force component parallel to
the direction of motion is ``recognized'' as the gradient of a random
potential, whereas the perpendicular components can {\em not} be
distinguished from a true random force, since the particle does not
explore these directions.  Therefore $\overline{u_x^2(t)} \approx
\frac {\Delta_0} {\eta^2 v^2 \xi^{d}}$ saturates for large times,
whereas $\overline{u_y^2(t)} \approx \frac {\Delta_0} {\eta^2 v
\xi^{d+1}} |t|$ grows without bounds, like under the influence of
thermal noise. This implies that the shaking temperature $T_{\rm sh}
\sim \frac{\Delta_0}{v}$ is associated with the {\em perpendicular}
displacement components.

\begin{figure}
\epsfxsize=0.9 \linewidth \epsfbox{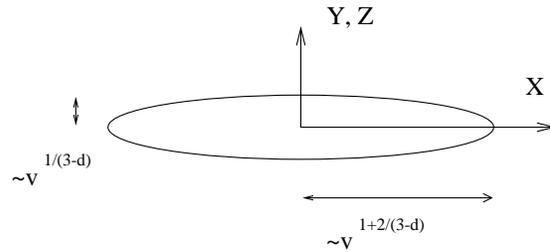} \narrowtext
\caption{Shape of the Larkin domain in the drifting structure. The
  domain is much longer in directions parralel ($x$) than
  perpendicular ($y,z$) to ${\bf v}$. Its anisotropy increases with
  increasing velocity, see Eq. (\ref{larkin}).}
\label{fig.larkin}
\end{figure}

The second feature of anisotropy, a more rapid growth of the relative
displacements $w({\bf r})$ in the direction perpendicular to the
motion (i.e. for ${\bf r}\perp {\bf v}$), is related to the size and
shape of the dynamic Larkin domain, in which the pinning forces act
coherently on the elastic medium. Since $u_x$ has much weaker
fluctuations than $u_y$ the correlation lengths are determined by the
fluctuations of $u_y$ alone and are found to be
\begin{equation}
y_c = a \left(\frac {c \xi^{d+3} \eta v}{a^2 \Delta_0} \right)^
  {\frac 1{3-d}}, \quad
  x_c = \eta v y_c^2/c ~.
\end{equation}
For weak disorder they are finite only in dimensions $d \leq 3$ ($y_c$
still depends logarithmically on $v$ in $d=3$) and increase for large
velocities much faster parallel than perpendicular to the velocity
(see Fig. \ref{fig.larkin}).

The next important question of the stability of the lattice with
respect to plastic relative displacements of vortices can be captured
by a phenomenological Lindemann criterion, that examines the
fluctuations in the relative distance of neighboring vortices (bond
length). Vortices neighboring in a direction parallel to ${\bf v}$
have much weaker fluctuations in their relative position than
neighbors in perpendicular directions. When the relative fluctuations
of certain bonds exceed a certain fraction of the vortex spacing,
these bonds are expected to be broken by topological defects. We find
that the bonds in directions {\em perpendicular} to the velocity have
the strongest fluctuations.  This consideration therefore supports the
suggestion that above a certain critical value of the velocity the VL
moves coherently like a solid and the topological order of the lattice
may be preserved despite of the roughness of the lattice in $d \leq
3$. Below this critical velocity the motion is plastic and vortices
may move in decoupled channels. It is essentially the anisotropy of
the Larkin domain that provides decoupling of flowing vortex channels.

\begin{figure}
  \epsfxsize=0.9 \linewidth \epsfbox{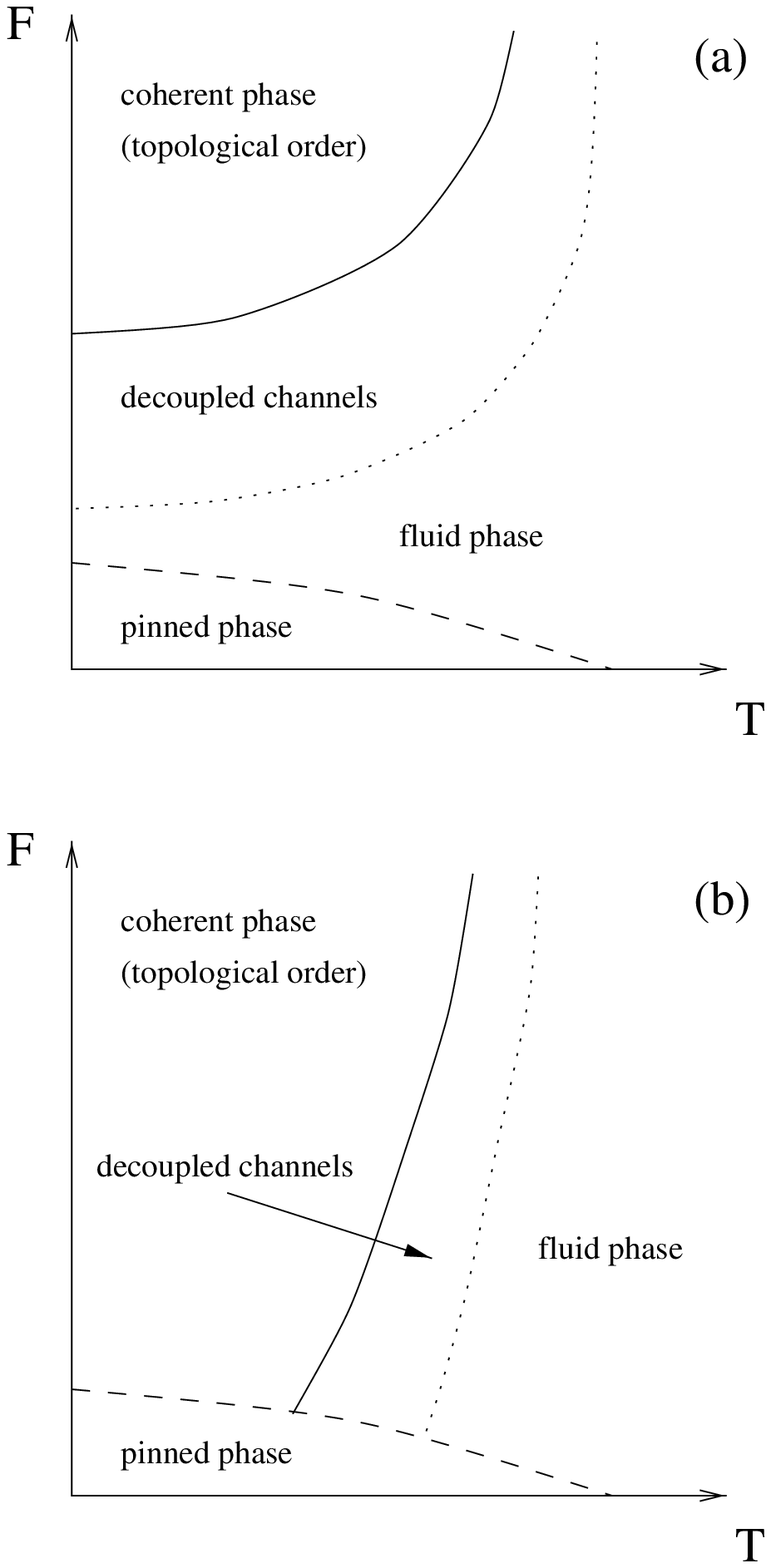} \narrowtext
  \caption{Topology of the generic phase diagram of driven periodic
  media (a) for strong disorder and (b) for weak disorder. At largest
  velocities the medium moves coherently. At the solid line a melting
  occurs into an incoherent phase with massive plasticity. This phase
  can display smectic properties and decay at even smaller velocities
  (dotted line) into a fluid-like phase. At even smaller velocities
  (dashed line) the creep phase with metastable states is reached.} 
\label{fig.phases}
\end{figure}

Fig. \ref{fig.phases} summarizes our view of the dynamic phase diagram
for the case of sufficiently strong disorder.  Starting from highest
drives we expect a {\it coherent} motion of the topologically ordered
phase.  Upon decreasing the driving force the fluctuations of the
bonds between the neighboring vortices cause a massive production of
topological defects at the transition from coherent to incoherent
motion marked by the solid line. This line corresponds to the freezing
transition of \cite{KV94}.  The question concerning the nature of the
plastically moving phase still remains.  Our analysis suggests that
there is a tendency to channel formation, but at this point we cannot
conclude whether these channels remain stable upon a further
decrease of the applied force, and therefore dynamic melting describes
transition from the moving quasi-crystal to moving smectic
\cite{BMR97}, or directly into the fluid like phase.  The possible
transition between the smectic and fluid-like phases is denoted by the
dotted line.  The lower strip below the critical current (the dashed
line) corresponds to the pinned state where the system moves via
thermally activated jumps between meta-stable states.

For weak disorder a dynamic transition from the coherently pinned
phase to a coherently moving phase is possible without passing through
a plastic regime. Plasticity occurs only at sufficiently high
temperature and for small enough velocities. Since the anisotropy of
the system decreases with decreasing velocity, the width of the
smectic regime shrinks in that direction.

In fact it also remains a fundamental open question to what extent the
creep regime can be considered as coherent in the sense that the
topological order persists up to the largest length scales. The
successful description of this regime by collective pinning requires
only the typical distance between free topological defects to increase
faster than the size of the largest effective barriers for decreasing
creep velocity. In principle it is possible that only at strictly zero
velocity the coherence of the lattice is restored.

The paper is organized as follows.  In Sec. \ref{sec.model} we specify
the model under consideration. In Sec. \ref{sec.gen} the general
perturbative approach is established in a dynamical formalism. A
scheme for the systematic extraction of coarse-grained parameters that
describe the physics at large scales is presented. These parameters
are evaluated in Sec.  \ref{sec.eval} and lead to our conclusions in
Sec \ref{sec.conc}.  The complexity of the problem requires a
compactified notation that is summarized in Appendix \ref{append.not}.
Intermediate steps of our calculations are sketched in Appendix
\ref{append.seco}.

\section{Model for driven vortex lattices}
\label{sec.model}

To be specific we introduce a model of a $d$-dimensional vortex
lattice. The most common realizations are vortex lines in a
three-dimensional superconductor ($d_\parallel=1$ and $d_\perp=2$),
point vortices in a film ($d_\parallel=0$ and $d_\perp=2$), or vortex
lines confined to a plane ($d_\parallel=1$ and $d_\perp=1$).  We use a
unifying description by considering vortices as
$d_\parallel$-dimensional objects that can be displaced in $d_\perp$
directions within a $d=d_\parallel+d_\perp$-dimensional
superconductor.

\begin{figure}
\epsfxsize=0.9 \linewidth \epsfbox{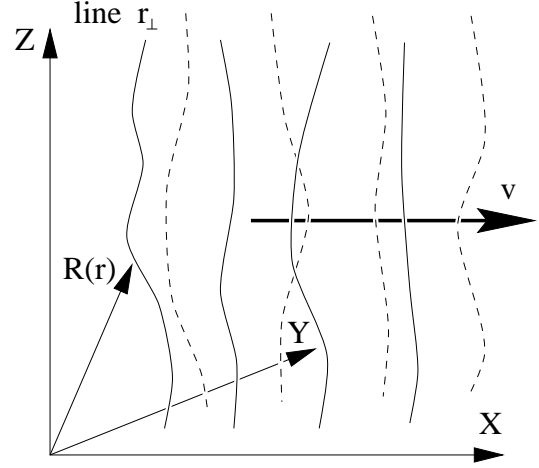} \narrowtext
\caption{Geometry of the vortex lattice in $d=1+2$. Vortices carry a
  fixed label ${\bf r}_\perp$ and their fluctuating position is ${\bf
  R}$. The $x$-axis is chosen parallel to the average velocity ${\bf
  v}$.}
\label{fig.geom}
\end{figure}

To every individual vortex we assign a fixed label ${\bf r}_\perp$
that coincides with its position in a perfectly ordered lattice. The
coordinates along the magnetic field are denoted by ${\bf
r}_\parallel$.  The actual position of vortex ${\bf r}_\perp$ at time
$t$ is denoted by ${\bf R}({\bf r},t)$, where ${\bf r}=({\bf
r}_\perp,{\bf r}_\parallel)$. For a three-dimensional vortex line
lattice ${\bf r}_\perp$ is a vector in the $(x,y)$-plane and ${\bf
r}_\parallel={\bf R}_\parallel$ represents the $z$-coordinate, see
Fig.  \ref{fig.geom}.

We consider a sliding state where vortices move with the average
velocity ${\bf v}$. Then ${\bf r}$ is viewed as the undistorted vortex
position in a comoving frame, whereas ${\bf R}$ is the actual position
in a laboratory frame. To parameterize the fluctuations of the vortex
lattice, we define vortex displacements as
\begin{equation}
  \label{def.u}
  {\bf u}({\bf  r},t)={\bf R}({\bf r},t) - {\bf r} - {\bf v}t ~.
\end{equation}
The proper choice of the perfect lattice positions and of ${\bf v}$
guarantees that these displacements always vanish upon averaging over
thermal fluctuations and disorder.

We restrict ourselves to the elastic lattice, where the topology of
the vortices is fixed and their interactions can be treated in the
harmonic approximation. The dynamics of the vortex lattice is governed
by the over-damped Langevin equation
\begin{eqnarray}
  \label{eqmo}
  \eta \dot {\bf u}({\bf r},t)&=&{\bf Du}({\bf r},t)
  + {\bf F}-\eta {\bf v}
  +{\bbox \xi}({\bf r},t) \nonumber \\
  &&+{\bf f} ({\bf r}+{\bf v}t + {\bf u}({\bf r},t)) 
\end{eqnarray}
with the Bardeen-Stephen friction coefficient $\eta$, the elastic
force ${\bf D}{\bf u}\sim c {\bbox \partial}^2 {\bf u}$ to be
specified below, a driving force ${\bf F}$, a thermal noise ${\bbox
\xi}({\bf r},t)$, and a pinning force ${\bf f}({\bf R})=-{\bbox
\partial} V({\bf R})$. Both the thermal and pinning forces are
supposed to have a Gaussian distribution with zero average and
correlations
\begin{mathletters}
  \label{def.corr}
  \begin{eqnarray}
    \langle \xi_\alpha({\bf r}_1,t_1) \xi_\beta({\bf r}_2,t_2) \rangle
    &=& \vartheta  \delta_{\alpha \beta} \delta({\bf
      r}_{12})\delta(t_{12}) ~,\\
    \vartheta &:=&  2 a^{d_\perp} \eta T ~, \\
    \overline {V({\bf R}_1) V({\bf R}_2)} &=& \Delta ({\bf R}_{12}) ~.
  \end{eqnarray}
\end{mathletters}
To make our formulas comprehensive and transparent, we introduce a
shortened notation, where e.g. ${\bf r}_{12}:={\bf r}_1-{\bf r}_2$ and
$t_{12}:=t_1-t_2$ (see also Appendix \ref{append.not} for
definiteness).  Greek indices represent components in the $d_\perp$
directions of ${\bf r}_\perp$.

\subsection{Action formulation}

The main difficulty in solving Eq. (\ref{eqmo}) is the highly
non-linear dependence of the pinning force on the displacements. We
will treat this non-linearity by a perturbative expansion in $1/v$.  A
convenient way to explore dynamics is the standard field-theoretical
representation of Martin, Siggia, and Rose \cite{MSR73,BJW76}.  In
this formalism the partition function for the out-of-equilibrium
system is defined as
\begin{eqnarray}
  \label{def.Z}
  Z&:=&\int {\cal D}_\Lambda [u,{\tilde u}] \ e^{-{\cal A}} ~,
  \nonumber \\
  {\cal D}_\Lambda [u,{\tilde u}]&:=& \prod_{t, {\bf q}}^{(\Lambda)}
  d^{d_\perp} {\tilde u}({\bf q},t) d^{d_\perp} u({\bf q},t) ~,
\end{eqnarray}
where the path integral is restricted to modes $q \leq \Lambda$ with
the cutoff $\Lambda$ \cite{Lambda}. This scale can be related to the
coherence length as ``diameter'' of the vortices.  The auxiliary
response field ${\tilde {\bf u}}$ is introduced in addition to the
displacement field ${\bf u}$.

To every possible configuration of the fields (including their
time-dependence) a statistical weight $e^{-{\cal A}}$ is assigned with
an action ${\cal A}={\cal A}[u,{\tilde u}]$. The sum over all weights
is normalized to unity and is independent of the random pinning
potential.  Therefore a disorder averaging can be performed
straightforwardly, which produces a translation-invariant effective
field theory.  We decompose the resulting action into the ``pure'' and
the ``pinning'' part
\begin{eqnarray}
  \label{def.action}
  {\cal A} &=& {\cal A}^{\rm pure} + {\cal A}^{\rm pin} ~,
  \nonumber \\
  {\cal A}^{\rm pure} &=& \int_1
  \left\{ \frac {\vartheta}2  {\tilde {\bf u}}_1 \cdot {\tilde {\bf
        u}}_1 + i {\tilde {\bf u}}_1 \cdot
    \left[\eta \dot {\bf u}_1 - ({\bf Du})_1 -
      {\bf F} + \eta {\bf v} \right] \right \} ~,
  \nonumber \\
  {\cal A}^{\rm pin} &=& \frac 12 \int_{12}
  {\tilde {\bf u}}_1 \cdot {\bbox \Phi}({\bf R}_{12}) \cdot
  {\tilde {\bf u}}_2 ~.
\end{eqnarray}
We have introduced further abbreviations in Eq. (\ref{def.action})
(see Appendix \ref{append.not}): an integer index $i$ stands for
$({\bf r}_i,t_i)$, ${\bf u}_i \equiv {\bf u}({\bf r}_i,t_i)$, ${\bf
R}_{12}\equiv {\bf r}_{12} + {\bf v} t_{12} + {\bf u}_{12}$, and the
scalar product includes all $d$ space components. $\int_i$ represents
a short-hand notation for an integration over $t_i$, an integration
over ${\bf r}_{\parallel i}$ and a summation over the vortex labels
${\bf r}_{\perp i}$. The latter summation includes the factor
$a^{d_\perp}$ representing the volume per one vortex (for the usual
vortex lattice with two displacement components $a^2=\Phi_0/B$ with
the flux quantum $\Phi_0$ and the magnetic induction $B$) such that
the sum would become an integral in the continuum limit $a \to 0$. It
is important to stress that although the discreteness of the lattice
does not manifest itself in the notation, the {\em discrete nature} of
the system is completely accounted for in our description.  The
pinning part contains the pinning force correlator ${\bbox \Phi}$,
which is related to the pinning potential correlator $\Delta$ by
\begin{equation}
  \label{def.Phi.gen}
  {\bbox \Phi}({\bf R})\equiv \Phi_{\alpha \beta}({\bf R}):=
  \overline{ f_\alpha({\bf R})  f_\beta({\bf 0})}=
  -\partial_\alpha \partial_\beta \Delta({\bf R}) ~.
\end{equation}
By a replacement $\Phi({\bf R}_{12}) \to \Phi({\bf r}_{12}, {\bf
R}_{12})$ the present theory can be generalized straightforwardly to
other problems like that of interfaces dynamics or random $XY$-models
\cite{jK95,GS85,S95}.

The discreteness of the vortex lattice enforces a periodicity in
Fourier space
\begin{equation}
  \label{def.FT.u}
  {\bf u}({\bf q}+{\bf Q},\omega) = {\bf u}({\bf q},\omega) :=
  \int_{t {\bf r}}
  e^{ i \omega t - i {\bf q} {\bf r}} \ {\bf u} ({\bf r},t)
\end{equation}
with reciprocal lattice vectors (RLV) ${\bf Q}$.

\subsection{Pure part}

The elastic interaction of vortices in the harmonic approximation is
most conveniently represented in Fourier space, where the pure part of
the action acquires the form
\begin{equation}
  \label{A.f}
  {\cal A}^{\rm pure} = \int_{\omega {\bf q}}
  \left\{ \frac 12 {\tilde {\bf u}^\dagger} \cdot  
  {\tilde{\bbox  \Gamma}} \cdot {\tilde {\bf u}} + 
  i {\tilde {\bf u}^\dagger} \cdot {\bbox \Gamma} \cdot {\bf u} \right\} ~.
\end{equation}
The symbol ${\bf u} \equiv {\bf u}({\bf q},\omega)$, ${\tilde {\bf
u}^\dagger} \equiv {\tilde {\bf u}} (-{\bf q},-\omega)$ (the dagger
stands for transposition of components and complex conjugation of
Fourier-transformed quantities), and the ${\bf q}$-integration runs
only over the first Brillouin-zone of the lattice. Note that one can
always choose ${\tilde{\bbox \Gamma}}^\dagger = {\tilde{\bbox
\Gamma}}$. In Eq. (\ref{A.f}) we have dropped the terms linear in the
response field. Since the average velocity is defined by the condition
that the average displacement has to vanish, these terms actually have
to cancel each other in the absence of disorder. This implies ${\bf
F}=\eta {\bf v}$.

It is convenient to write the propagators of the pure action in the
normal mode representation. In the $d_\perp=2$-dimensional case these
modes are longitudinal ($L$) or transverse ($T$) with respect to ${\bf
q}_\perp$. Using the projectors
\begin{equation}
  \label{def.project}
      P_{L,\alpha\beta} ({\bf q}):=\frac {q_\alpha q_\beta}{q_\perp^2}
      ~,\quad
    P_{T,\alpha\beta} ({\bf q}):=
    \delta_{\alpha \beta}- P_{L,\alpha\beta}({\bf q})
\end{equation}
we have ($p=L,T$)
\begin{equation}
  \label{def.decompose}
     {\bf \tilde\Gamma} ({\bf q},\omega)
    := \sum_p \tilde\Gamma_p({\bf q},\omega) {\bf P}_p ({\bf q})
\end{equation}
and a similar expression for $ {\bf \Gamma}$, where
\begin{mathletters}
  \label{props}
  \begin{eqnarray}
    \tilde \Gamma_p ({\bf q},\omega)&:=& \vartheta ~,
    \\
    \Gamma_p ({\bf q},\omega) &:=& -i \eta \omega + D_p({\bf q}) ~.
  \end{eqnarray}
\end{mathletters}
The elastic dispersion relations read
\begin{mathletters}
  \begin{eqnarray}
    \label{def.LT}
    D_{L}({\bf q}) &=& c_{11} q_\perp^2 + c_{44} q_\parallel^2 ~,\\
    D_{T}({\bf q}) &=& c_{66} q_\perp^2 + c_{44} q_\parallel^2 ~,
  \end{eqnarray}
\end{mathletters}
where the elastic constants for compression $c_{11}$, shear $c_{66}$,
and tilt $c_{44}$ of the vortex lattice can have additional implicit
dependences on ${\bf q}$ \cite{Bla+94}.

Within the partition sum (\ref{def.Z}) response- and correlation-
functions are defined as
\begin{mathletters}
  \label{def.GC}
  \begin{eqnarray}
    {\bf G}_{12} \equiv G_{\alpha \beta}({\bf r}_{12},t_{12})&:=&
    \langle u_\alpha({\bf r}_1,t_1) i
    {\tilde u}_\beta({\bf r}_2, t_2) \rangle ~,\\
    {\bf C}_{12} \equiv C_{\alpha \beta}({\bf r}_{12},t_{12})&:=&
    \langle u_\alpha({\bf r}_1,t_1) u_\beta({\bf r}_2,t_2) \rangle ~.
  \end{eqnarray}
\end{mathletters}
These averages are to be performed with the action (\ref{def.action})
that has already been averaged over disorder. Therefore no further
disorder-averaging is required in (\ref{def.GC}) and these two-point
quantities depend only on time and space differences. $G_{\alpha
\beta}({\bf r}_{12},t_{12})=\delta \langle u_\alpha({\bf r}_1,t_1)
\rangle / \delta \xi_\beta({\bf r}_2,t_2)$ describes the response of
the displacement ${\bf u}$ to the thermal noise ${\bf \xi}$ or some
additional external force acting on the vortices and is therefore
causal, i.e. ${\bf G}({\bf r},t)=0$ for $t \leq0$ \cite{Ito}.

In the sequel the average squared displacement of vortices
\begin{equation}
  C_{0, \alpha \beta}:=
  \langle u_\alpha({\bf r},t) u_\beta ({\bf r},t) \rangle
\end{equation}
will play a particular role, as well as the difference correlation
\begin{eqnarray}
  W_{\alpha \beta}({\bf r}_{12},t_{12}):= \frac 12 \langle
  [u_{1\alpha} - u_{2\alpha}] [u_{1\beta} - u_{2\beta}] \rangle
  \nonumber \\
  = C_{0 \alpha \beta}-\frac 12 C_{\alpha \beta}({\bf r}_{12},t_{12})
  -\frac 12 C_{\alpha \beta}({\bf r}_{21},t_{21}) ~.
\end{eqnarray}

For the general pure action (\ref{A.f}) the response and correlation
functions are related to the propagators by
\begin{mathletters}
  \label{GC.ai}
  \begin{eqnarray}
    {\bf G}({\bf q},\omega) &=& {\bbox \Gamma}^{-1}({\bf q},\omega) ~, \\
    {\bf C}({\bf q},\omega) &=& {\bf G}({\bf q},\omega) \cdot
    {\tilde{\bbox \Gamma}}({\bf q},\omega) \cdot
    {{\bf G}^\dagger} ({\bf q},\omega) ~.
  \end{eqnarray}
\end{mathletters}
The last equation implies ${\bf C}^\dagger = {\bf C}$ in agreement
with the definition (\ref{def.GC}b).

In the special case (\ref{props}) we have explicitly
\begin{mathletters}
  \label{gen.corr}
  \begin{eqnarray}
    G_p({\bf q},\omega) &=&
    \frac 1 {\Gamma_p ({\bf q},\omega)}
    =  \frac 1{ -i \eta \omega + D_p({\bf q}) } ~, \\
    C_p({\bf q},\omega) &=& \frac{{\tilde \Gamma}_p ({\bf q}, \omega) }
    {|\Gamma_p ({\bf q},\omega)|^{2}}
    =  \frac {\vartheta} {\eta^2 \omega^2 + D_p^2({\bf q})} ~.
  \end{eqnarray}
\end{mathletters}
The decomposition (\ref{def.decompose}) into normal modes holds also
for these matrices.

\section {Perturbation theory: General scheme}
\label{sec.gen}

In this Section we perform a perturbative coarse-graining for the
dynamics of the vortex lattice. We find an effective dynamics for the
fields with wave-vectors $q \leq \Lambda^<$ below a reduced cutoff
$\Lambda^<<\Lambda$ by averaging over all modes with wave-vectors
$\Lambda^< < q \leq \Lambda$. This averaging is performed by
integrating out these modes in the partition sum (\ref{def.Z}). Due to
the high non-linearity of the pinning part of the action this
integration cannot be performed exactly. We restrict our perturbative
analysis to the first and second order in the pinning action.

As usual we separate the modes with wave vectors below and above the
new cutoff,
\begin{mathletters}
  \begin{eqnarray}
    \label{def.split.u}
    {\bf u} &=& {\bf u}^< + {\bf u}^> ~, \\
    {\bf u}^<({\bf r},t) &=& \int_{{\bf q}\leq \Lambda^<} e^{i {\bf q}
      \cdot {\bf r}} \ {\bf u}({\bf q},t) ~, \\
    {\bf u}^>({\bf r},t) &=& \int_{\Lambda^< <{\bf q} \leq \Lambda}
    e^{i {\bf q} \cdot {\bf r}} \ {\bf u}({\bf q},t) ~.
  \end{eqnarray}
\end{mathletters}
The response field ${\tilde {\bf u}}$ is treated analogously. The
integration over the modes ${\bf u}^>$ and ${\tilde {\bf u}}^>$ leads
then to an effective action for the modes ``$<$''.  The effective
action can be represented via a cumulant expansion which is
\begin{mathletters}
  \label{A.cumul}
  \begin{eqnarray}
    {\cal A}^{\rm eff}&=&{\cal A}+{\cal A}^{(1)}+{\cal A}^{(2)}+\dots
    ~, \\
    {\cal A}^{(1)}[{\tilde u}^<,u^<]&=&
    \langle {\cal A}^{\rm pin}[{\tilde u}^<+{\tilde u}^>,u^<+u^>] \rangle^>
    \nonumber  \\ &&
    -{\cal A}^{\rm pin}[{\tilde u}^<,u^<] ~, \\
    {\cal A}^{(2)}[{\tilde u}^<,u^<]&=& - \frac 12
    \langle {\cal A}^{\rm pin}[{\tilde u}^<+{\tilde u}^>,u^<+u^>],
    \nonumber \\
    && {\cal A}^{\rm pin}[{\tilde u}^<+{\tilde u}^>,u^<+u^>] \rangle^>_{c}
  \end{eqnarray}
\end{mathletters}
to the second order perturbation theory in ${\cal A}^{\rm pin}$. The
averaging is performed over the modes ${\bf u}^>$ and ${\tilde {\bf u}}^>$
weighted with the pure action ${\cal A}^{\rm pure}[{\tilde u}^>,u^>]$. We
denote the second cumulant by $\langle {\cal A}, {\cal A}
\rangle_c:=\langle {\cal A}^2 \rangle -\langle {\cal A} \rangle^2$.

Now we turn to a detailed analysis of the above corrections and derive
new couplings in the effective action for the large-scale modes.

\subsection{First order}

Using Fourier-transforms of the disorder correlator
\begin{equation}
  \label{A^i}
  {\cal A}^{\rm pin}=
  \frac 12 \int_{12{\bf k}}
  {\tilde {\bf u}}_1 \cdot {\bbox \Phi}({\bf k}) \cdot {\tilde {\bf u}}_2
  \ e^{i {\bf k} \cdot {\bf R}_{12}} 
\end{equation}
we shift all the dependences on the displacements to the exponential.
Applying Wick's theorem to the first order correction given in Eq.
(\ref{A.cumul}b) we get
\begin{eqnarray}
  \label{A^1}
  {\cal A}^{(1)}[{\tilde u}^<,u^<]= \int_{12 {\bf k}}
  e^{i{\bf k} \cdot {\bf R}^<_{12} -
    {\bf k} \cdot {\bf W}^>_{12} \cdot  {\bf k}}
  \nonumber \\  \times
  \left\{ \frac 12 {\tilde {\bf u}}_1^< \cdot
    {\bbox \Phi}({\bf k}) \cdot {\tilde {\bf u}}_2^<
    + {\tilde {\bf u}}_1^< \cdot {\bbox \Phi}({\bf k})
    \cdot {{\bf G}^>_{21}}^\dagger
    \cdot {\bf k} \right\} ~.
\end{eqnarray}
The terms $\sim {\bf G}^>_{12} {\bf G}^>_{21}$ vanish due to
causality.  In addition terms $\sim {\bf G}^>_{11}$ etc. vanish in the
Ito calculus. As a result all terms that appear in the correction to
the action contain at least one response field. The superscripts
``$\gtrless$'' in Eq. (\ref{A^1}) stands to remind that ${\bf
R}^<_{12} \equiv {\bf r}_{12}+{\bf v} t_{12}+{\bf u}_{12}^<$ and that
${\bf G}^>$ and ${\bf W}^>$ arises from the averaging over modes
``$>$''. To keep the notation simple, we will hereafter drop these
superscripts.

The obtained correction to the action has a more complicated
functional dependence on the fields than the original action ${\cal
  A}$. From ${\cal A}^{(1)}$ we extract not only corrections to the
parameters of the original ${\cal A}$, but also new couplings like a
Kardar-Parisi-Zhang (KPZ) nonlinearity \cite{KPZ86}.

\subsubsection{Friction force}

The conventional approach to the description of driven disordered
dynamics is to fix the external drive ${\bf F}$ and evaluate the
resulting drift velocity ${\bf v} \equiv {\bf v}({\bf F})$ as a
response thereof. However, since in the present formulation ${\bf F}$
appears only at one position in the action, whereas ${\bf v}$ appears
at many positions, we find it more convenient technically to treat the
average velocity as given and to derive the force ${\bf F} \equiv {\bf
F}({\bf v})$ necessary to maintain this velocity. The fact that ${\bf
v}$ is indeed the average vortex velocity is expressed by the
condition $\langle {\bf u}({\bf q=0},\omega=0) \rangle=0$. In the
presence of disorder this requires that the {\em effective} action for
large-scale modes has no terms linear in ${\tilde {\bf u}}$, i.e. no
terms of order ${\cal O}({\tilde u}^1 u^0)$ in the fields. In the
absence of disorder this immediately yields that the driving force is
compensated by the friction force ${\bf F}^{\rm fr}=\eta {\bf v}$.

The first-order correction to the friction force is extracted from
order ${\cal O}({\tilde u}^1 u^0)$ of (\ref{A^1}):
\begin{equation}
  \label{A^1_1}
  {\cal A}^{(1)}_{{\tilde u}^1 u^0} =  
  \int_{12 \bf k}
  {\tilde {\bf u}}_1 \cdot {\bbox \Phi}({\bf k}) \cdot 
  {{\bf G}^\dagger}_{21} \cdot {\bf k} 
  \ e^{i{\bf k} \cdot ({\bf r}_{12} + {\bf v} t_{12}) 
    -{\bf k} \cdot {\bf W}_{12} \cdot {\bf k}} ~.
\end{equation}
Comparing this contribution to the term of order ${\cal O}({\tilde
  u}^1 u^0)$ in the original action we identify the first order
perturbative correction to the pinning force
\begin{eqnarray}
  \label{F^1}
  F^{{\rm fr}(1)}_\alpha &=& -i \int_{t\bf r k} 
  e^{i{\bf k} \cdot ({\bf r} + {\bf v} t) - 
    {\bf k} \cdot {\bf W}({\bf r},t) \cdot {\bf k}}
  \nonumber \\ && \times
  \Phi_{\alpha \beta} ({\bf k}) G_{\gamma \beta}({\bf r},t) k_\gamma ~.
\end{eqnarray}
Consequently the transport characteristics of the superconductor is
given by ${\bf F}^{\rm fr}({\bf v})=\eta {\bf v} + {\bf F}^{{\rm
fr}(1)}({\bf v})$. Since ${\bf F}^{(1)}$ is in general a non-linear
function of ${\bf v}$, the characteristics is no longer linear over
the whole current range.

\subsubsection{Friction coefficient, elastic dispersion}

The friction coefficient $\eta$ and the elastic dispersion relations
$D$ parameterize the propagator ${\tilde{\bbox \Gamma}}$ in the
original pure action (\ref{A.f}) in the order ${\cal O}({\tilde u}^1
u^1)$. Therefore, corrections to these parameters are extracted from
(\ref{A^1}) in the same order,
\begin{eqnarray}
  \label{A^1_11}
  {\cal A}^{(1)}_{{\tilde u}^1 u^1}&=&
  \int_{12 \bf k} 
  e^{i{\bf k} \cdot ({\bf r}_{12} +{\bf v} t_{12}) 
    -{\bf k} \cdot {\bf W}_{12} \cdot {\bf k} }
  \nonumber \\ && \times
  [{\tilde {\bf u}}_1 \cdot {\bbox \Phi}({\bf k}) \cdot 
  {{\bf G}^\dagger}_{21} \cdot {\bf k}]
  [i{\bf k} \cdot ({\bf u}_1 -{\bf u}_2)] ~.
\end{eqnarray}
The integral kernel now has a finite width in $({\bf r}_{12},t_{12})$.
However, this width is smaller than that of the response function,
which decays on a characteristic scale $1/\Lambda^<$ in space and on a
scale $\eta/D(\Lambda^<) $ in time.

Aiming at the physics at scales much larger than the width of the
kernel we consider the displacement fields as nearly constant and
approximate
\begin{eqnarray}
  \label{appr.diff}
  {\bf u}_1-{\bf u}_2 &\approx& t_{12} \partial_t {\bf u}_1 + 
  r_{12a} \partial_a {\bf u}_1 
  \nonumber \\ &&
  -\frac 12 r_{12a} r_{12b} 
  \partial_a \partial_b {\bf u}_1 + \dots ~.
\end{eqnarray}
(Latin indices also include directions parallel to the vortex lines.)
Inserting (\ref{appr.diff}) in (\ref{A^1_11}) we find
\begin{equation}
  \label{A^1_11.approx}
  {\cal A}^{(1)}_{{\tilde u}^1 u^1} =
  \int_{\omega \bf q}  i {\tilde {\bf u}^\dagger}  
  \cdot {\bbox \Gamma}^{(1)} \cdot {\bf u} ~,
\end{equation}
where the correction to the response propagator can be written as
\begin{equation}
  \label{Gt.1}
  \Gamma_{\alpha \beta}^{(1)}=-i \eta_{\alpha \beta}^{(1)} \omega + 
  D_{\alpha \beta}^{(1)}({\bf q})
\end{equation}
with an elastic dispersion
\begin{equation}
  \label{D^1}
  D^{(1)}_{\alpha \beta}({\bf q}) = 
  - i \chi^{(1)}_{\alpha \beta a} q_a
  +\kappa^{(1)}_{\alpha \beta ab} q_a q_b ~.
\end{equation}
Therein new coefficients $\chi$ appear. They describe forces arising
from a direct coupling to lattice stresses. In addition, the
coarse-graining generates elastic constants $\kappa$ with reduced
symmetries compared to the original constants.

The correction to the friction coefficient is explicitly
\begin{eqnarray}
  \label{eta^1}
  \eta^{(1)}_{\alpha \beta} &=&
  \int_{t\bf rk}
  e^{i{\bf k} \cdot ({\bf r}+{\bf v}t) 
    - {\bf k} \cdot {\bf W}({\bf r},t) \cdot {\bf k} }
  \nonumber \\
  && \times
  t \Phi_{\alpha \gamma}({\bf k}) 
  G_{\delta \gamma}({\bf r},t) k_\delta k_\beta ~.
\end{eqnarray}
In general this tensor is non-diagonal and gives rise to Hall effects.
The stress coefficients are
\begin{eqnarray}
  \label{chi^1}
  \chi^{(1)}_{\alpha \beta a} &=& 
  - \int_{t \bf rk}  
  e^{i{\bf k} \cdot ({\bf r}+{\bf v}t)
    - {\bf k} \cdot {\bf W}({\bf r},t) \cdot {\bf k} }
  \nonumber \\
  && \times
  r_a \Phi_{\alpha \gamma}({\bf k}) 
  G_{\delta \gamma}({\bf r},t) k_\delta k_\beta
\end{eqnarray}
and the elastic couplings are corrected by
\begin{eqnarray}
  \label{kappa^1}
  \kappa^{(1)}_{\alpha \beta ab} &=& 
  \frac 12 \int_{t \bf rk}  
  e^{i{\bf k} \cdot ({\bf r}+{\bf v}t)
    - {\bf k} \cdot {\bf W}({\bf r},t) \cdot {\bf k} } 
  \nonumber \\
  && \times
  r_a r_b \Phi_{\alpha \gamma}({\bf k}) 
  G_{\delta \gamma}({\bf r},t) k_\delta k_\beta ~.
\end{eqnarray}

Comparing Eqs. (\ref{F^1}) and (\ref{eta^1}) one finds
straightforwardly the following useful relation:
\begin{equation}
  \label{fric.1} 
  \eta_{\alpha \beta}^{(1)}({\bf v}) = \frac {\partial
  F_\alpha^{\rm fr(1)} ({\bf v})}{\partial v_\beta} ~.
\end{equation}
In the first order perturbation theory the friction coefficient
coincides with the differential resistivity.

\subsubsection{The KPZ term}

Action (\ref{A^1}) contains a further contribution in order ${\cal
O}({\tilde u}^1 u^2)$:
\begin{eqnarray}
  \label{A^1_12}
  {\cal A}^{(1)}_{{\tilde u}^1 u^2}&=& - \frac 12 \int_{12 \bf k}
  e^{i{\bf k} \cdot ({\bf r}_{12}+{\bf v} t_{12}) 
    -{\bf k} \cdot {\bf W}_{12} \cdot {\bf k} } 
  \nonumber \\ && \times
  [{\tilde {\bf u}}_1 \cdot {\bbox \Phi}({\bf k}) \cdot 
  {{\bf G}^\dagger}_{21} \cdot {\bf k}]
  [{\bf k} \cdot ({\bf u}_1-{\bf u}_2)]^2 ~.
\end{eqnarray}
As before, we use the approximation (\ref{appr.diff}) which leads
to (we omit other terms that are less relevant on large scales) 
\begin{equation}
  \label{A^1_12.approx}
  {\cal A}^{(1)}_{{\tilde u} u u}=
  \int_{t \bf r} i {\tilde u}_\alpha ({\bf r},t) 
  \left\{ - \frac 12 \lambda^{(1)}_{\alpha\beta\gamma ab}
    [\partial_a u_\beta({\bf r},t)] [\partial_b u_\gamma({\bf r},t)] \right\}
\end{equation}
with 
\begin{eqnarray}
  \label{lambda^1}
  \lambda^{(1)}_{\alpha\beta\gamma ab} &=& 
  - i \int_{t \bf rk} 
  e^{i{\bf k} \cdot ({\bf r}+{\bf v}t) 
    - {\bf k} \cdot {\bf W}({\bf r},t) \cdot {\bf k} }
  \nonumber \\ && \times
  r_a r_b k_\beta k_\gamma k_\epsilon
  \Phi_{\alpha \delta}({\bf k}) G_{\epsilon \delta}({\bf r},t) ~.
\end{eqnarray}
This means that in the driven state disorder induces a KPZ term.

\subsubsection{Disorder correlator}

So far we have not considered ${\cal O}({\tilde u}^2)$ of (\ref{A^1}):
\begin{equation}
  \label{A^1_2}
  {\cal A}^{(1)}_{{\tilde u}^2}= \frac 12 \int_{12 \bf k} 
  [{\tilde {\bf u}}_1 \cdot {\bbox \Phi}({\bf k}) \cdot {\tilde {\bf u}}_2 ]
  e^{i{\bf k} \cdot {\bf R}_{12}}
  [e^{ -{\bf k} \cdot {\bf W}_{12} \cdot {\bf k} }-1] ~.
\end{equation}
Comparing the functional form of this action to (\ref{A^i}) we
identify a correction ${\bbox \Phi}^{(1)}({\bf k})$ to the disorder
correlator from the {\em persistent} part of the kernel, i.e. the part
that is present also for $|t_{12}| \to \infty$. Using $\lim_{|t_{12}|
\to \infty} {\bf W}({\bf r}_{12},t_{12})={\bf C}_0$ we find
\begin{equation}
  \label{Phi^1}
  {\bbox \Phi}^{(1)}({\bf k}) = {\bbox \Phi}({\bf k})
  [e^{-{\bf k} \cdot {\bf C}_0 \cdot {\bf k}}-1] ~.
\end{equation}
Remarkably, ${\bbox \Phi}^{(1)}$ is independent of velocity and
vanishes in a perfectly ordered lattice with ${\bf C}_0=0$. In the
general case with finite ${\bf C}_0$ the correction represents a
smearing-out of disorder by the vortex fluctuations.

\subsubsection{Temperature}

The remaining {\em non-persistent} part of (\ref{A^1_2}), which is not
taken into account by the disorder correction, is
\begin{eqnarray}
  \label{A^1_2.np}
  {\cal A}^{(1) {\rm np}}_{{\tilde u}^2}&=& \frac 12 \int_{12 \bf k} 
  e^{i{\bf k}\cdot {\bf R}_{12}}
  \left[e^{ -{\bf k} \cdot {\bf W}_{12} \cdot {\bf k} }
    - e^{-{\bf k} \cdot {\bf C}_0 \cdot {\bf k}} \right]
  \nonumber \\ && \times 
  [{\tilde {\bf u}}_1 \cdot {\bbox \Phi}({\bf k}) \cdot 
  {\tilde {\bf u}}_2 ] ~.
\end{eqnarray}
Now this integrand vanishes for $|t_{12}| \to \infty$ and is also
local in ${\bf r}_{12}$. Assuming again that the width of this kernel
is small compared to the scales of variation of ${\bf u}$ and ${\tilde
  {\bf u}}$, one may neglect ${\bf u}_1-{\bf u}_2$ (as zeroth-order of
approximation (\ref{appr.diff})) and approximate
\begin{equation}
  \label{A^1_11.np.approx}
  {\cal A}^{(1) {\rm np}}_{{\tilde u}^2}= \frac 12 \int_{t \bf r} 
  {\tilde {\bf u}}({\bf r},t) \cdot {\bbox \vartheta}^{(1)} \cdot 
  {\tilde {\bf u}}({\bf r},t) ~.
\end{equation}
Herein
\begin{equation}
  \label{vartehata^1}
  \vartheta^{(1)}_{\alpha \beta}  
  =\int_{t \bf rk} 
  e^{i{\bf k} \cdot ({\bf r}+{\bf v}t)}
  \left[e^{ -{\bf k} \cdot {\bf W}({\bf r},t) \cdot {\bf k} }
    - e^{-{\bf k} \cdot {\bf C}_0 \cdot {\bf k}} \right] 
  \Phi_{\alpha \beta}({\bf k})
\end{equation}
is the correlator of the effective thermal noise (non-persistent
shaking forces). Note that $\vartheta^{(1)}_{\alpha \beta} =0$ for a
perfectly ordered lattice with ${\bf C}_0={\bf 0}$.

\subsection{Second order}

The second-order correction ${\cal A}^{(2)}$ to the action has to be
calculated according to Eq. (\ref{A.cumul}c).  The result contains a
large number of terms, and the full expression is not displayed here.
One can easily see, however, that ${\cal A}^{(2)}$ contains terms of
orders ${\cal O}({\tilde u}^1)$, ${\cal O}({\tilde u}^2)$, and ${\cal
O}({\tilde u}^3)$.  From ${\cal O}({\tilde u}^1)$ corrections to the
pinning force and to ${\bbox \Gamma}$ can be extracted. From ${\cal
O}({\tilde u}^2)$ one derives corrections to the disorder correlator
and to ${\tilde{\bbox \Gamma}}$. Eventually, new types of couplings
appear in ${\cal O}({\tilde u}^3)$ that represent higher-order
cumulants of the disorder, i.e. deviations from a Gaussian
distribution.

The subsequent analysis is restricted to the evaluation of the
correction to the disorder correlator, in order to demonstrate the
presence of effective random forces in the coarse-grained equation of
motion. The second order correction to the disorder correlator
${\bbox \Phi}^{(2)}$ is again obtained by identifying the contributions to
${\cal O}({\tilde u}^2)$ in ${\cal A}^{(2)}$ that represent persistent
(quenched) forces in the laboratory frame in contrast to temporarily
fluctuating forces that contribute to the effective thermal noise.

Due to the complexity of the involved expressions the technical
details of this procedure are deferred to Appendix \ref{append.seco}.

\subsubsection{Random force}

Now we determine the coarse-grained disorder correlator. The
second-order correction is extracted from ${\cal A}^{(2)}$ in ${\cal
O}({\tilde u}^2)$. We identify ${\bbox \Phi}^{(2)}({\bf k})$ as the
kernel where slowly varying displacements and response fields enter in
exactly the same combination as they appeared in the original pinning
action (\ref{A^i}). Therefore ${\bbox \Phi}^{(2)}({\bf k})$ represents
the correlator of forces that are {\em stationary in the laboratory
frame}. The force experienced by a vortex moving in the laboratory
frame is nevertheless fluctuating in time. 

The calculation of this correlator is performed in Appendix
\ref{append.seco} and leads to the somewhat involved expression for
${\bbox \Phi}^{(2)}({\bf k})$ given in Eq.  (\ref{Phi^2}). From the
large scale behavior (${\bf k} \to 0$)
\begin{equation}
  \label{Phi.split}
  \Phi_{\alpha \beta}({\bf k})
  =\phi_{\alpha \beta}+ \phi_{\alpha \beta \gamma} k_\gamma 
  + \frac 12  \phi_{\alpha \beta \gamma \delta} k_\gamma k_\delta + \dots
\end{equation}
one identifies the {\em random force} correlator $\phi_{\alpha
  \beta}$. This contribution as well as the second term in
(\ref{Phi.split}) emerge only in the driven state {\it and} in the
presence of disorder. The bare random potential contributes only to the
coefficient $\phi_{\alpha \beta \gamma \delta}=2\delta_{\alpha \gamma}
\delta_{\beta \delta} \Delta({\bf k}=0)$.

Eq. (\ref{Phi^2}) gives explicitly
\begin{eqnarray}
  \label{phi}
  \phi_{\alpha \beta}^{(2)} =  \int_{12 \bf k} 
  e^{- {\bf k} \cdot [{\bf W}({\bf r}_1,t_1)+{\bf W}({\bf r}_2,t_2)] 
    \cdot {\bf k} 
    + i {\bf k} \cdot {\bf R}^{(0)}_1 }
  \nonumber \\ 
  \times k_\sigma G_{\sigma \gamma}(-{\bf r}_2,-t_2) 
  k_\tau G_{\tau \delta} ({\bf r}_1,t_1)
  \nonumber \\ 
  \times \left\{ 
    \Phi_{\alpha \beta}({\bf k}) \Phi_{\gamma \delta}({\bf k}) 
    e^{i {\bf k} \cdot {\bf R}^{(0)}_2}
    -
    \Phi_{\alpha \gamma}({\bf k}) \Phi_{\beta \delta}({\bf k}) 
    e^{-i {\bf k} \cdot {\bf R}^{(0)}_2}
  \right\}
\end{eqnarray}
with ${\bf R}^{(0)}_i:={\bf r}_i+{\bf v}t_i$.

\subsubsection{Other terms}

One can also extract coefficients $\phi_{\alpha \beta \gamma}$ and
$\phi_{\alpha \beta \gamma \delta}$ of the force correlator from
(\ref{Phi^2}).  However on large length scales the corresponding terms
in the action are less relevant than the random force, and therefore
we do not present them here.
 
A number of other terms appear in the second order of perturbation
theory introducing in particular new types of disorder.  For example,
taking into account the gradient term in the expansion
(\ref{appr.diff}) one finds a random correction to the amplitude of
the KPZ nonlinearity as suggested by Krug \cite{jK95}.  Also the
second-order corrections to ${\bf F}$ and to the propagators
appear. Again, these corrections have a complicated form and are not
given here. For weak disorder the second-order corrections are
expected to be small compared to the first-order corrections.

\subsection{Fluctuation-Dissipation Theorem}

In the absence of disorder the system obeys the
Fluctuation-Dissipation Theorem (FDT) \cite{DH75}
\begin{equation}
  \label{FDT}
  -2 i \eta \omega {\tilde \Gamma}_{\alpha \beta} ({\bf q},\omega) = 
  \vartheta [\Gamma_{\alpha \beta} ({\bf q},\omega) -
  \Gamma_{\beta \alpha} (-{\bf q},-\omega) ] ~,
\end{equation}
which implies (for $t \neq 0$)
\begin{eqnarray}
  \label{FDT.0}
  2 \eta \partial_t W_{\alpha \beta}({\bf r},t)&=&\vartheta 
  [G_{\alpha \beta}({\bf r},t)+G_{\beta \alpha}({\bf r},t)
  \nonumber \\ &&
  -G_{\alpha \beta}(-{\bf r},-t)-G_{\beta \alpha}(-{\bf r},-t)] ~.
\end{eqnarray}
The FDT holds even in the driven state due to the Galilei-invariance
(the response- and correlation functions under consideration are
defined in the comoving frame).

In the presence of disorder the validity of the FDT would require
\begin{eqnarray}
  \label{FDT.1}
  -2 i \eta \omega {\tilde \Gamma}_{\alpha \beta}^{(1)} ({\bf
    q},\omega) = \vartheta 
  [\Gamma_{\alpha \beta}^{(1)} ({\bf q},\omega) -
  \Gamma_{\beta \alpha}^{(1)} (-{\bf q},-\omega) ] ~.
\end{eqnarray}
At zero velocity this relation is satisfied even in the presence of
disorder, because the terms $\chi$ vanish and $\eta
\vartheta^{(1)}_{\alpha \beta} = \vartheta \eta^{(1)}_{\alpha \beta}$.
However, in the {\it driven} dirty system the FDT is violated.  The
most obvious reason is the presence of the stress couplings generated
by disorder in the driven system, also one sees immediately that
$\vartheta^{(1)}$ and $\eta^{(1)}$ no longer satisfy the
aforementioned relation.

\section{Evaluation of perturbation theory}
\label{sec.eval}

In the previous section we have derived the effective equation of
motion (\ref{eqmo^*}) for the coarse-grained displacement. The
renormalized parameters (with superscript $^*$) are composed of the
original values plus perturbative corrections,
\begin{eqnarray*}
  \eta_{\alpha \beta}^*&=&\eta \delta_{\alpha \beta}+\eta_{\alpha
    \beta}^{(1)} ~, 
  \\
  D_{\alpha \beta}^*&=& D_{\alpha \beta} + D_{\alpha \beta}^{(1)} ~,
  \\
  F_\alpha^{{\rm fr}*} &=& \eta v_\alpha + F_\alpha^{{\rm fr}(1)} ~, 
  \\
  \lambda_{\alpha \beta \gamma ab}^* &=& 
  \lambda_{\alpha \beta \gamma  ab}^{(1)} ~, 
  \\
  \langle \xi_\alpha^*({\bf r}_1,t_1)\xi_\beta^*({\bf r}_2,t_2) \rangle
  &=&\vartheta_{\alpha \beta}^* \delta({\bf r}_{12})\delta(t_{12}) ~,
  \\
  \vartheta_{\alpha \beta}^* &=& \vartheta \delta_{\alpha \beta} 
  + \vartheta_{\alpha \beta}^{(1)} ~,
  \\
  \overline{f_\alpha^*({\bf R}_1)f_\beta^*({\bf R}_2)} &=& 
  \Phi_{\alpha \beta}^*({\bf R}_{12}) ~,
  \\
  \Phi_{\alpha \beta}^*&=&\Phi_{\alpha \beta} + \Phi_{\alpha \beta}^{(1)}
  +\Phi_{\alpha \beta}^{(2)} ~.
\end{eqnarray*}
Now we evaluate the various couplings that appeared upon the
coarse-graining procedure. We start with the first-order perturbative
corrections in subsections \ref{eval.rp} - \ref{eval.KPZ} and then
derive the random-force term as the second-order correction in
subsection \ref{eval.rf}.

To simplify the evaluation of the general perturbative results we
consider disorder with an isotropic correlation length $\xi$
\begin{equation}
  \label{Delta.spec}
  \Delta({\bf k})= \Delta_0 e^{- \frac 12 \xi^2 k^2} ~,
\end{equation}
where we assume $\xi \ll a$, which is typical for superconductors. In
addition, we assume the elastic constants to be uniform,
i.e. $D_{\alpha \beta}({\bf q})=\delta_{\alpha \beta} c q^2$ whenever
explicit expressions involving elastic constants are given.  This will
not change our results qualitatively.

As shown by Schmid and Hauger \cite{SH73} the lattice prefers to move
along the principal symmetry axes. Hereafter we restrict the analysis
to the situation where the velocity is parallel to a high-symmetry
direction of the lattice, which we choose to be the $x$-axis.

\subsection{Random potential}
\label{eval.rp}

We start with the discussion of the coarse-grained disorder
correlator. The first-order correction (\ref{Phi^1}) preserves the
random-potential nature of the original disorder. Coarse-graining
smears out the correlation length of the disorder over the typical
vortex displacement since $\Phi_{\alpha \beta}({\bf k})+
\Phi^{(1)}_{\alpha \beta}({\bf k})=e^{-{\bf k} \cdot {\bf C}_0 \cdot
{\bf k}} \Phi_{\alpha \beta}({\bf k})=k_\alpha k_\beta \Delta_0
e^{-{\bf k} \cdot {\bf C}_0 \cdot {\bf k} - \frac 12 \xi^2 k^2}$. This
means that the disorder correlation lengths is described by the matrix
\begin{equation}
  \label{def.Xi}
  \Xi_{ab}:= \xi^2 \delta_{ab} + C_{0ab} ~.
\end{equation}
The correction $C_{0ab}$ diverges in dimensions $d \leq 2$ for finite
temperatures, if the lower cutoff $\Lambda^<$ is sent to zero. In this
case it is therefore possible that weak disorder is irrelevant for the
large-scale properties of the vortex lattice. As long as $\Lambda^<>0$
finite this correction has only a quantitative effect in all
dimensions.

Focusing on $d>2$ we will consider in what follows mainly the case of
zero temperature. In most perturbative expressions the exponential
factors involving matrices ${\bf C}_0$ or ${\bf W}$ can be ignored
since they always enter in a combination with the disorder correlator
and modify the correlation lengths only quantitatively. The only
exception is provided by the correction to the temperature.

\subsection{Temperature}

We evaluate the effective temperature from Eq. (\ref{vartehata^1}),
where we had observed already that the correction vanishes for the
perfectly ordered lattice at $T=0$, where ${\bf W}={\bf C}_0={\bf
0}$. At finite temperatures one expects a positive correction, since
in general ${\bf W} < {\bf C}_0$ and the difference between the
exponentials in Eq. (\ref{vartehata^1}) will no longer
vanish. Considering low temperatures, one may linearize the
exponentials and obtains
\begin{equation}
\vartheta^{(1)}_{\alpha \beta} \approx 
\int_{\bf k} k_\alpha k_\beta k_\perp^2 \Delta({\bf k}) 
\frac {\vartheta}{\eta^2 v^2 k_x^2 + c^2 q^4} ~,
\end{equation}
where we implicitly decompose ${\bf k}={\bf Q}+{\bf q}$ into a
reciprocal lattice vector (RLV) ${\bf Q}$ and a vector ${\bf q}$
within the first Brillouin zone. With a random-potential disorder
(\ref{Delta.spec}) one immediately finds in the limit of large
velocities $v \gg \xi c/\eta a^2$
\begin{mathletters}
  \label{vartheta.xx}
  \begin{eqnarray}
    \vartheta^{(1)}_{xx} &\approx& \frac{\Delta_0}{ \xi^{2+d} \eta^2 v^2} 
    \vartheta ~,   \\
    \vartheta^{(1)}_{yy} &\approx& \frac{a^2 \Delta_0}{\xi^{3+d}c \eta v}  
    \vartheta  ~ {\rm for} ~ d>3 ~, \\
    \vartheta^{(1)}_{yy} &\approx& \frac{a^2 \Delta_0}{\xi^{3+d}c \eta v}
    (a \Lambda^<)^{d-3} \vartheta   ~ {\rm for} ~ d<3 ~.
  \end{eqnarray}
\end{mathletters}
We have dropped numerical factors of order unity.  For large
velocities the result for $\vartheta^{(1)}_{xx}$ is independent of the
elastic constants and finite for $\Lambda^< \to 0$.

However the integrand of $\vartheta^{(1)}_{yy}$ has poles near ${\bf k}
\approx {\bf Q}$ giving rise to the divergence in $d \leq 3$ (the
divergence is logarithmic in $d=3$) when the coarse-graining cutoff
$\Lambda^<$ is sent to zero. This divergence is the first indication
that perturbation theory is insufficient for the true large-scale
description of the drifting vortex lattice even for arbitrarily large
drive.

Since the correlator $\vartheta^{(1)}_{yy}$ decays even for finite
$\Lambda^<$ with a smaller power in $v$ than $\vartheta^{(1)}_{xx}$,
the effective noise is stronger in the directions perpendicular to the
velocity. Since only $\vartheta^{(1)}_{yy}$ exhibits this divergence,
there is a fundamental difference between vortices with $d_\perp > 1$
displacement components and CDW with $d_\perp = 1$.

This divergence therefore does {\em not} imply that fluctuations,
which have on small scales an amplitude proportional to temperature in
the pure case, would now diverge. Since this divergence of the
effective temperature arises from the large-scale response of the
elastic medium to the pinning force, it suggests only that on large
scales the medium is much more rough than in the pure case. However,
the roughness does not depend on the effective temperature only, but
also on the effective friction coefficients and elastic constants,
which we thus have to evaluate before we can address the roughness.

\subsection{Friction force}

At $T=0$ and uniform elasticity the evaluation of Eq. (\ref{F^1})
yields
\begin{equation}
\label{F.x}
  F_\alpha^{{\rm fr} (1)}= 
  \eta v \int_{\bf k} k_x k_\alpha k_\perp^2 \Delta({\bf k})
  \frac{1}  {\eta^2 v^2 k_x^2+ c^2 q^4} ~,
\end{equation}
The relative correction to the friction force closely related to the
relative correction (\ref{vartheta.xx}) of the effective temperature.
We find explicitly for $v \gg \xi c/\eta a^2$
\begin{equation}
  \label{F.xx}
  F_x^{{\rm fr} (1)} \approx \frac {\Delta_0}{\xi^{d+2} \eta v} ~,
\end{equation}
where purely numerical prefactors of order unity have been dropped.
This expression is independent of the elastic constant for large
velocities. In this regime vortices respond dynamically like
individual particles. The typical force scale
$F_0^2=\Delta_0/\xi^{d+2}$ is set by the spatial average of the
pinning force.

Eq. (\ref{F.xx}) applies only to velocities parallel to a principal
lattice axis. For other directions the force and velocity will no
longer be parallel to each other \cite{SH73}: the velocity will
deviate from the force in the direction of the closest principal
lattice axis. Thus disorder induces a Hall-effect. Instead of
evaluating the friction force for arbitrary directions, we will turn
to the friction coefficients that describe the same effect in a
differential form. There the presence of the Hall-effect shows up as
anisotropy of the coefficient matrix.

At small velocities Eq. (\ref{F.x}) gives Ohmic behavior, $F_x^{{\rm
fr} (1)} \sim v$, only for $d > 4$. In this case the poles of the
integrand at ${\bf k}={\bf Q}$ with ${\bf Q} \cdot {\bf v}=0$ are
integrable. This is no longer true in $d \leq 4$ and the perturbation
theory gives sub-Ohmic transport $F_x^{{\rm fr} (1)} \sim
v^{(d-2)/2}$. In particular in $d=2$ the effective friction force is
finite for small velocities \cite{SH73}. The sub-Ohmic behavior of the
effective friction force reflects the fact that the vortex lattice
forms a glass {\em below four dimensions and at zero velocity}
\cite{aiL70}.

\subsection{Friction coefficients}

In order to examine to what extent glassy features persist in the
driven state at finite velocities, it is necessary to examine the
friction coefficients $\eta_{\alpha \beta}$. Since these coefficients
describe the dynamical response of the driven vortex lattice, the
glassy features, which are in general associated with divergent
relaxation times, must manifest themselves as divergences in the
friction coefficient.

Combining (\ref{F.x}) with (\ref{fric.1}) we find 
\begin{equation}
  \label{eta.x}
  \eta_{\alpha \beta}^{(1)} = - \eta \int_{\bf k} 
  k_\alpha k_\beta k_\perp^2 \Delta({\bf k}) 
  \frac{\eta^2 v^2 k_x^2 - c^2 q^4} {[\eta^2 v^2 k_x^2+ c^2 q^4]^2} ~. 
\end{equation}
Note that only the diagonal components of the friction coefficient do
not vanish due to reflection symmetries. In the limit of large
velocities parallel to a principal lattice axis ($x$-axis) one finds:
\begin{mathletters}
  \label{eta.xx}
  \begin{eqnarray}
    \eta^{(1)}_{xx} & \approx & 
    - \frac{\Delta_0}{\xi^{d+2} \eta^2 v^2} \eta
    \\
    \eta^{(1)}_{yy} & \approx &  
    \frac{a\Delta_0}{\xi^{d+3} \eta^2 v^2} \eta
  \end{eqnarray}
\end{mathletters}
The main contributions to $\eta^{(1)}_{yy}$ come from the vicinity
${\bf k} \approx {\bf Q}$ of RLV with ${\bf Q} \cdot {\bf v}=0$,
whereas for $\eta^{(1)}_{xx}$ all terms ${\bf Q} \cdot {\bf v} \neq 0$
contribute. We find that $|\eta^{(1)}_{yy}| \gg |\eta^{(1)}_{xx}|$
since we have evaluated Eq. (\ref{eta.x}) in the limit $\xi \ll
a$. Otherwise both corrections would be of the same order of
magnitude.

Since (\ref{F.xx}) has already been specified to velocities along a
principal lattice axis, only $\eta^{(1)}_{xx}$ can be derived from
there using the differential relation (\ref{fric.1}). Both friction
coefficients are again independent of elastic constants.

In their pioneering work Schmid and Hauger \cite{SH73} have discussed
a discontinuity in the relation between the transverse force and
velocity [see their Eq. (27) and discussion thereafter]. However, as
they state, this discontinuity is an artifact due to a neglect of
vectors ${\bf q}$. Such a discontinuity, if real, should have appeared
as a divergence in $\eta^{(1)}_{yy}$, which is actually {\em absent}
\cite{note.eta}.

Glassy features in the dynamical response normal to the velocity,
suggested by Giamarchi and Le Doussal \cite{GL96}, should manifest
themselves as a divergence of the friction coefficient $\eta_{yy}$. At
large drive this divergence can emerge only in higher orders of
perturbation theory.

At zero velocity a divergence $\eta_{\alpha \beta}^{(1)} \sim
{\Lambda^<}^{d-4}$ is apparent for $d \leq 4$. It arises from the
poles of the integrand in (\ref{eta.x}) at ${\bf k}={\bf Q}$ and
implies $\eta_{\alpha \beta}^{(1)} \sim v^{(d-4)/2} \delta_{\alpha
\beta}$ for $\Lambda^< =0$. This again signals the glassiness at ${\bf
v=}0$ for $d \leq 4$.

\subsection{Stress coefficients}

For $T=0$ and uniform elasticity Eq. (\ref{chi^1}) reduces to
\begin{equation}
  \label{chi.x}
  \chi^{(1)}_{\alpha \beta a} = 
  4 \eta v \int_{\bf k} k_x k_\alpha k_\beta k_\perp^2 \Delta({\bf k})  
  \frac{c^2 q_a q^2} {[\eta^2 v^2 k_x^2 + c^2 q^4]^2} ~.
\end{equation}
We again restrict our consideration to the case where the velocity is
parallel to a principal lattice axis. For several combinations of
indices the integrand is odd under an inversion $k_a \to
-k_a$. Therefore we find in particular
\begin{equation}
  \chi^{(1)}_{\alpha \beta z} =
  \chi^{(1)}_{yxx} =
  \chi^{(1)}_{xyx} =
  \chi^{(1)}_{xxy} = 
  \chi^{(1)}_{yyy} = 0 ~.
\end{equation}
(We would like to remind that Greek indices run only over the
directions perpendicular to the vortex lines, i.e. $x$ and $y$ in the
usual 3D configuration, whereas the Latin indices also include
directions parallel to the vortex lines, i.e. $z$ in the usual 3D
configuration.) But for finite velocity there are also non-vanishing
components which decay in the limit $v \to \infty$ like
\begin{mathletters}
  \label{chi.xx}
  \begin{eqnarray}
    \chi^{(1)}_{xxx} &\approx& \frac {\Delta_0 c^2}{\xi^{d+1} a^3 \eta^3
      v^3} ~, 
    \\
    \chi^{(1)}_{xyy} &=& \chi^{(1)}_{yxy} \approx \frac {\Delta_0 c}
    {\xi^{d+1} a^2 \eta^2 v^2} ~,
    \\
    \chi^{(1)}_{yyx} &\approx&  \frac {\Delta_0 c}
    {\xi^{d+3} \eta^2 v^2} ~.
  \end{eqnarray}
\end{mathletters}
We have assumed $\xi \ll a$. Then $\chi^{(1)}_{yyx}$ is the largest
among these coefficients.

\begin{figure}
\epsfxsize=0.9 \linewidth
\epsfbox{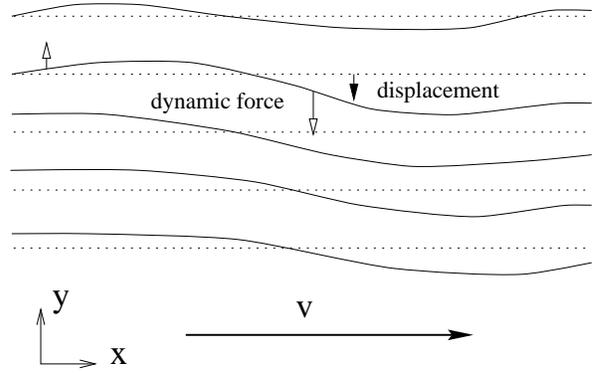}
\narrowtext
\caption{Schematic representation of a moving vortex
  configuration. The distance between the solid wiggly line (position
  of vortex in perfect lattice) and the dotted straight line (actual
  vortex position) represents the component $u_y$ of the displacements
  field.  The force dynamically generated force $f_y \sim \chi_{yyx}
  \partial_x u_y$, visualized by empty arrows at two representative
  positions, makes a vortices acquire the displacement of the
  preceding vortex.}
\label{fig.chi}
\end{figure}

The stress coefficients have a simple physical meaning. They express
the tendency of vortices to form a homogeneously moving system. In
particular the coefficients $\chi_{xxx}>0$ and $\chi_{yyx}>0$ imply
that a vortex experiences a dynamical force that makes it follow the
motion (``footsteps'') of the precedent vortex, see Fig.
\ref{fig.chi}. These couplings favor the formation of channels.

The stress coefficients are generated only in a non-equilibrium driven
state and in the presence of disorder. At small velocity and for
$\Lambda^<=0$ they vanish $\sim v$ only in dimensions $d>4$, where the
integral in (\ref{chi.x}) is finite. For $d \leq 4$ the stress
coefficients vanish sub-linearly. Lowest-order perturbation theory
gives e.g.  $\chi_{xxx} \sim v^{(d-2)/2}$. Thus these coefficients
diverge in $d<2$ only.

\subsection{Elastic constants}

Following the same scheme a correction to the elastic constants
\begin{eqnarray}
  \label{kappa.x}
  \kappa_{\alpha \beta ab}^{(1)}&=&
  - c \int_{\bf k} k_\alpha k_\beta k_\perp^2 \Delta({\bf k}) 
  \Big\{
  \delta_{ab} \frac{\eta^2 v^2 k_x^2  - c^2 q^4}
  {[\eta^2 v^2 k_x^2 + c^2 q^4]^2}
  \nonumber \\
  && -4 c^2 q_a q_b q^2 \frac{3 \eta^2 v^2 k_x^2 - c^2 q^4}
  {[\eta^2 v^2 k_x^2 + c^2 q^4]^3}\Big\} 
\end{eqnarray}
can be obtained from Eq. (\ref{kappa^1}).

For symmetry reasons again all coefficients $\kappa_{\alpha \beta
ab}^{(1)}$ vanish where indices different from $x$ appear an odd
number of times. Nevertheless $\kappa_{\alpha \beta ab}^{(1)}$ will no
longer be proportional to $\delta_{\alpha \beta} \delta_{ab}$ and the
correction reduces the symmetry of the original elastic interaction.

The first term in (\ref{kappa.x}) represents a correction of the
elastic constant proportional to the correction of the friction
coefficients (\ref{eta.x}). This contribution is finite for all
dimensions and of the order $\sim v^{-2}$ for large $v$. The second
term in (\ref{kappa.x}) can be shown to be finite and of order $\sim
v^{-4}$.  Therefore, at large velocities, it can be neglected in
comparison to the first term.

We find thus in the leading order $\sim v^{-2}$ 
\begin{mathletters}
\label{kappa.xx}
\begin{eqnarray}
\kappa_{xxab}^{(1)}&\approx& 
- \frac{\Delta_0}{\xi^{d+2} \eta^2 v^2} \delta_{ab} \ c ~, \\
\kappa_{yyab}^{(1)}&\approx& 
\frac{a \Delta_0}{\xi^{d+3} \eta^2 v^2} \delta_{ab} \ c ~.
\end{eqnarray}
\end{mathletters}
Hence the elastic stiffness of the lattice for displacements {\em
parallel} to the velocity is {\em reduced}, whereas the stiffness for
displacements {\em perpendicular} to the velocity is {\em
increased}. The latter effect can be interpreted as a tendency to form
channels and an increased energy cost for transverse displacements.

For small velocities the corrections to the elastic constants are
found to diverge $\kappa^{(1)} \sim v^{(d-4)/2}$ like the friction
coefficients.

\subsection{KPZ nonlinearity}
\label{eval.KPZ}

As before we obtain from  Eq. (\ref{lambda^1})
\begin{eqnarray}
  \label{lambda.x}
  \lambda_{\alpha \beta \gamma ab}^{(1)}&=&
  4 c \eta v \int_{\bf k} k_x k_\alpha k_\beta k_\gamma  
  k_\perp^2 \Delta({\bf k}) 
  \Big\{
  \frac{\delta_{ab} c q^2} {[\eta^2 v^2 k_x^2 + c^2 q^4]^2}
  \nonumber \\
  && + 2 c q_a q_b \frac{\eta^2 v^2 k_x^2 - 3 c^2 q^4}
  {[\eta^2 v^2 k_x^2 + c^2 q^4]^3}\Big\} ~.
\end{eqnarray}
Many terms vanish due to symmetry. All coefficients that are related
by permutations among Greek or among Latin indices are identical.

We find in particular
\begin{mathletters}
  \label{lambda.xx}
  \begin{eqnarray}
    \lambda_{xxxab}^{(1)}&\approx& 
    \frac{c^2 \Delta_0}{a^2 \xi^{d+2} \eta^3  v^3} \delta_{ab} ~,
    \\
    \lambda_{xyyab}^{(1)}&\approx& 
    \frac{c \Delta_0}{\xi^{d+3} \eta^2  v^2} \delta_{ab} ~, 
    \\
    \lambda_{yyyyx}^{(1)}&\approx& 
    - \frac{c^2 \Delta_0}{a \xi^{d+3} \eta^3  v^3} ~.
  \end{eqnarray}
\end{mathletters}
All couplings assume finite values and decay at least $\sim v^{-2}$ in
the limit of large driving.

For CDW, which are included in our analysis by specializing all Greek
indices to $x$, Chen et al. \cite{CBFM96} have found
$\lambda_{xxxab}^{(1)} \sim v^{-1}$ in contrast to our result
(\ref{lambda.xx}a). Having no access to their derivation, we were not
able to pinpoint the origin of the disagreement.

In the limit of small velocities the KPZ terms coefficients vanish
$\sim v$ only as long as the integral in (\ref{lambda.x}) is finite,
i.e. for $d >6$. For $d<6$ the coefficients scale like $\lambda \sim
v^{(d-4)/2}$. One can see that these coefficients diverge even for
$d<4$!

\subsection{Random force}
\label{eval.rf}

At $T=0$ and uniform elasticity the evaluation of the random force
correlator Eq. (\ref{phi}) yields
\begin{equation}
\label{phi.x}
  \phi_{\alpha \beta}^{(2)}=2 \int_{\bf k} k_\alpha k_\beta k_\perp^4
  \Delta^2({\bf k}) \frac{\eta^2 v^2 k_x^2}
  {[\eta^2 v^2 k_x^2 + c^2 q^4]^2} ~.
\end{equation}
This expression was recently given in \cite{GL97} where it apparently
was found within a RG framework. Here we obtain it as a result of a
straightforward perturbation theory.

One easily calculates for  $v \gg \xi c/\eta a^2$ 
\begin{mathletters}
  \label{phi.xx}
  \begin{eqnarray}
    \phi^{(2)}_{xx} &\approx& \frac{\Delta_0^2}{ \xi^{4+d} \eta^2 v^2} ~,
    \\
    \phi^{(2)}_{yy} &\approx& \frac{a^2 \Delta_0^2}{\xi^{3+d}c \eta v}  
    ~ {\rm for} ~ d>3 ~, \\
    \phi^{(2)}_{yy} &\approx& \frac{a^2 \Delta_0^2}{\xi^{3+d}c \eta v}
    (a \Lambda^<)^{d-3}  ~ {\rm for} ~ d<3 ~.
  \end{eqnarray}
\end{mathletters}
For large velocities the result for $\phi^{(2)}_{xx}$ is independent
of the elastic constants and finite for $\Lambda^< \to 0$. This random
force is the analogon of the random mobility of driven interfaces with
phase disorder \cite{jK95}. A similar force for CDW was previously
obtained in \cite{CBFM96,BMR97} and predicted for vortex lattices also
in \cite{BMR97,GL97c}.

Comparing this result to the noise correlator (\ref{vartheta.xx}), we
find exactly the same type of divergences in the {\em transverse}
components in the limit $\Lambda^< \to 0$.

At small velocities the random force correlator (\ref{phi.x}) vanishes
like $v^{2}$ in $d > 8$ and $\sim v^{(d-4)/2}$ in lower dimensions
which again confirms the glassy nature at zero velocity in $d \leq
4$. The random-force component of the disorder correlator does not
always vanish in the limit $v \to 0$, it even diverges for $d<4$ in a
similar way like the KPZ coefficients.

\subsection{Roughening by disorder}

In the absence of disorder the displacements of the vortices are
isotropic and scale like $W_{xx}({\bf r},t) \sim W_{yy}({\bf r},t)
\sim w({\bf r},t) \sim b^{2 \zeta} w({\bf r}/b,t/b^z)$ with a
dynamical exponent $z=2$ and a thermal roughness exponent $\zeta=
(2-d)/2$.

The effect of disorder on the correlations can be estimated in the
most elementary approximation as follows: Assuming that the typical
displacements are ``small'' we might simply neglect them in the
argument of ${\bbox \Phi}$ in Eq. (\ref{def.action}). More precisely,
the validity of this approximation requires ${\bf u}_1-{\bf u}_2$ to
be small in comparison to ${\bf R}_{12}^{(0)}:={\bf r}_{12}+ {\bf
v}t_{12}$.

In this approximation the action is still bilinear in the fields and
of the same functional form (\ref{A.f}) as in the pure case. From
${\cal A}^{\rm pin}$ originates an additional contribution
\begin{equation}
  \label{Delta.Gt}
  \Delta {\tilde{\bbox \Gamma}}({\bf q},\omega)=
  \sum_{\bf Q} {\bbox \Phi} ({\bf k}) 
  \delta (\omega + {\bf k} \cdot {\bf v}) ~,
\end{equation}
which is the pinning force correlator as ``seen by the perfectly
ordered vortex lattice''. Here ${\bf k}$ is ${\bf k} \equiv {\bf
  Q}+{\bf q}$.

The large scale properties of the vortex lattice are in a good part
governed by the behavior of $\Delta {\tilde{\bbox \Gamma}}$ at small
${\bf q}$ and $\omega$. Note that there is an important difference
between the properties of components $\Delta \tilde \Gamma_{xx}({\bf
q},\omega)$ and $\Delta \tilde \Gamma_{yy}({\bf q},\omega)$.  It
follows straightforwardly from Eq. (\ref{Delta.Gt}) that the
correlator $\Phi_{\alpha \beta}({\bf k})$ contributes to the
asymptotic behavior not only at ${\bf k}=0$ but at all RLV ${\bf
k}={\bf Q}$ with ${\bf Q} \cdot {\bf v}=0$.  Therefore $\Delta \tilde
\Gamma_{yy}({\bf q},\omega)$ shows the same qualitative behavior at
small ${\bf q}$ and $\omega$ for both random potential (similar to
unrenormalized disorder) and random force case (appearing upon
coarse-graining). In both cases $\Delta \tilde \Gamma_{yy} \sim \delta
(\omega + {\bf q} \cdot {\bf v})$ with a finite prefactor for ${\bf
q}, \omega \to 0$.  In contrast, the behavior of the $\Delta \tilde
\Gamma_{xx}({\bf q},\omega)$ generated by the random potential is
qualitatively different from that of a random force. Namely, in the
random potential case $\Delta \tilde \Gamma_{xx}({\bf q},\omega)$
vanishes as $\Delta \tilde \Gamma_{xx}({\bf q},\omega) \sim q_x^2
\delta (\omega + {\bf q} \cdot {\bf v})$, whereas for the random force
case the $\delta$-function again has a finite prefactor. Thus the
random-force character of the coarse-grained disorder will change the
asymptotic behavior of those quantities which depend on $\tilde
\Gamma_{xx}$ but will leave intact those depending on $\tilde
\Gamma_{yy}$ only.

In the Gaussian approximation disorder does not modify the propagator
${\bbox \Gamma}$ and, according to Eq. (\ref{GC.ai}), leaves also the
response function unchanged. However, the additional contribution
(\ref{Delta.Gt}) to the propagator ${\tilde{\bbox \Gamma}}$ generates
an additional contribution
\begin{equation}
  \label{C.Gauss}
  \Delta C_{\alpha \beta}({\bf q},\omega)=
  \sum_{\bf Q} \frac{ \Phi_{\alpha \beta} ({\bf k})}
  {\eta^2 v^2 k_x^2 + c^2 q^4}
  \delta(\omega + k_x v) 
\end{equation}
to the correlation function. 

It is interesting to examine the dependence of correlations on space
and time. Since disorder is fixed in the laboratory frame, one might
expect that $\Delta C$ is also stationary in that frame, i.e.  $\Delta
C_{\alpha \beta}({\bf r},t)= C_{\alpha \beta}({\bf r}+{\bf v} t,0)$ or
$C_{\alpha \beta}({\bf q},\omega) \propto \delta(\omega + q_x v)$.  An
inspection of Eq.  (\ref{C.Gauss}) immediately reveals that only RLV
with $Q_x=0$ give such stationary contributions. All other
contributions, which reflect the discreteness of the vortex lattice in
the direction of the velocity are {\em not stationary}, neither in
the laboratory nor in the comoving frame.

Using the unrenormalized disorder $\Phi_{\alpha \beta} ({\bf k}) =
k_\alpha k_\beta \Delta_0 e^{-\frac 12 \xi^2 k^2}$ Eq. (\ref{C.Gauss})
implies (in $d \leq 3$) \cite{GL96}
\begin{mathletters}
  \label{W.Gauss}
  \begin{eqnarray}
    \Delta W_{xx}({\bf r},0) &\approx& \frac{\Delta_0}{\xi^d \eta^2
      v^2} ~ {\rm for} ~ r \gg a ~,\\
    \Delta W_{yy}({\bf r},0) &\approx& \frac{a^2 \Delta_0}
    {c \xi^{d+1} \eta v} \left( \frac {|y|} {a} \right)^{3-d} ~ 
    {\rm for}~ \eta v y^2 \gg c|x| ~,\\ 
    \Delta W_{yy}({\bf r},0) &\approx& \frac{a^2 \Delta_0}
    {c \xi^{d+1} \eta v} \left( \frac{c|x|}{a^2 \eta v}\right) ^{\frac{3-d}2} 
    ~ {\rm for} ~ \eta v y^2 \ll c|x| ~.
  \end{eqnarray}
\end{mathletters}
$\Delta W_{xx}$ has contributions from the vicinity of all RLV to
order $v^{-2}$. $\Delta W_{yy}$ is dominated on large scales by
contributions with RLV $Q_x=0$, whereas RLV with $Q_x \neq 0$ give
only finite contributions as to $\Delta W_{xx}$. As for $C$ also only
the contributions $Q_x=0$ are stationary in the laboratory frame. This
includes all contributions that roughen the VL on large scales. 

In this perturbative result the transverse displacement components
exhibit much stronger fluctuations than the longitudinal component. An
anisotropy emerges requiring thus a distinct scaling for the
displacement components parallel and perpendicular to the velocity and
also a distinct scaling for their dependence on coordinate distances
parallel and perpendicular to ${\bf v}$. While $W_{xx}$ does not
reveal a well-defined scaling behavior at large scales since finite
temperature and disorder give finite contributions (in $d>2$), the
scaling of $W_{yy}$ is dominated by the disorder contribution with
$W_{yy}(x,y,t) \sim b^{3-d} W_{yy}((x+vt)/b^2,y/b,0)$. The divergence
of $W_{yy}$ on large scales indicates that the Gaussian approximation
loses its validity at large scales, since the initial neglect of the
dependence of the force correlator on the displacements breaks
down. The characteristic length scales
\begin{mathletters}
  \label{larkin}
  \begin{eqnarray}
  y_c &=& a \left(\frac {c \xi^{d+3} \eta v}{a^2 \Delta_0} \right)^
  {\frac 1{3-d}} \\
  x_c &=& \eta v y_c^2/c
\end{eqnarray}
\end{mathletters}
that limit the validity range are obtained by
$w=W_{\alpha\alpha} =\xi^2$ (generalizing the static Larkin length)
have been introduced by Giamarchi and Le Doussal \cite{GL96}.

If one takes into account that a random-force is generated, which we
may approximate as $\Phi_{\alpha \beta}({\bf k}) \approx \phi_{\alpha
\beta}^{(2)} e^{-\frac 12 \xi^2 k^2}$ with the coefficients from Eq.
(\ref{phi.xx}) [an exponential decay for large ${\bf k}$ follows from
Eq. (\ref{Phi^2})], then also the $x$-component of the displacement
becomes rough in $d \leq 3$:
\begin{mathletters}
  \label{W.Gauss.2}
  \begin{eqnarray}
    \Delta W_{xx}({\bf r},0) &\approx& 
    \frac{a^2 \phi_{xx}}
    {c \xi^{d-1} \eta v} \left(\frac {|y|}{a} \right)^{3-d} ~ 
    {\rm for} ~ \eta v y^2 \gg c|x| ~,\\ 
    \Delta W_{xx}({\bf r},0) &\approx& 
    \frac{a^2 \phi_{xx}}
    {c \xi^{d-1} \eta v} \left( \frac{c|x|}{\eta va^2 }\right) ^{\frac{3-d}2} 
    ~ {\rm for} ~ \eta v y^2 \ll c|x|  ~.
  \end{eqnarray}
\end{mathletters}
However, the fluctuations of the transverse component are eventually
more pronounced since $\phi_{yy}$ and $\vartheta_{yy}$ diverge on
largest scales (for $y^{-1} \sim \Lambda^< \to 0$). This divergence is
logarithmic in $d=3$ and algebraic in $d<3$. Because of this
divergence renormalization effects are expected to modify the
roughness exponents found perturbatively. In addition, one would
expect naively that $\Delta W_{yy}$ increases on large scales even
faster than in Eq. (\ref{W.Gauss}). However, for consistency one
should take into account not only the one-loop corrections to
${\tilde{\bbox \Gamma}}$, but also to ${\bbox \Gamma}$.  In the static
case, where the Gaussian approximation yields a roughness exponent
$\zeta_0 =(4-d)/2$, the actual roughness is only logarithmic,
$\zeta=0$. There the disorder contribution to the force correlator
${\tilde{\bbox \Gamma}}$, that tends to increase the roughness, is
balanced by the contributions to ${\bbox \Gamma}$, notably an increase
to the elastic constants that increases the stiffness of the lattice.

For many disordered systems different approaches to treat a balance
between several diverging terms have been fruitful.  One possibility
are self-consistent approaches like that of Sompolinski and Zippelius
\cite{SZ} which treats the coupling of modes on a mean-field
level. This approach has been applied for example to spin glasses
\cite{SZ} and elastic manifolds \cite{em.mf}. However, since typically
the self-consistency takes into account only first-order corrections
${\cal A}^{(1)}$ to the action, it can produce only approximate values
of scaling exponents. In the present situation these approaches would
completely miss the physics arising from the divergence of
$\phi_{yy}$. It would therefore be necessary to extend this scheme to
the second-order corrections which make a solution of the
self-consistency equations even more involved. Another approach is
provided by the renormalization group (RG), that in principle can be
extended systematically to arbitrary perturbative order.

A consistent and reliable treatment of the large scale properties
requires the simultaneous handling of several complications: (i) The
anisotropies as discussed above start to interplay under RG
interaction with the anisotropies of the friction coefficient, stress
coefficients, elastic constants, and KPZ terms. (ii) Since disorder
roughens the VL, as seen already within the Gaussian approximation,
the generated KPZ terms are more relevant than the elastic couplings
according to scaling arguments. Therefore, a priori they also need to
be taken into account and are expected to modify the large scale
physics qualitatively as soon as the medium is rough. The relevance of
anisotropies in the KPZ terms in the absence of disorder has been
studied for {\em single} driven vortex lines by Ertas and Kardar
\cite{EK}, who find a variety of different physical regimes depending
on the anisotropy of elastic constants and KPZ coefficients, and by
Hwa \cite{tH92} for driven line liquids.  For CDW in $d=1,2$ Chen et
al.  \cite{CBFM96} find non-trivial scaling exponents (i.e.  differing
from the scaling found in the Gaussian approximation) due to the KPZ
terms also in the presence of disorder.

\section{Dislocations}

Our discussion has been restricted so far to the elastic
approximation, neglecting topological defects (like dislocations) in
the vortex lattice. Upon increasing velocity all effects of disorder
become weaker, and we expect also the length beyond which dislocations
become relevant to increase and even to diverge for temperatures below
the melting temperature of the pure system. Then the interesting
question arises: what kind of defects lead at smaller drift velocities
to a destruction of the coherence of motion, and what kind of order
can survive?

Balents, Marchetti, and Radzihovsky (BMR) \cite{BMR97} have proposed
the existence of a smectic phase where vortices are correlated over
much larger distances parallel to the drift velocity than
perpendicular to it.

We propose here a picture for the formation of a smectic phase within
a phenomenological approach based on a generalized Lindemann
criterion, examining the relative fluctuations in the distance of
neighboring vortices. Instead of addressing the topological defects
explicitly, we rather focus on their effect, namely the destruction of
the neighborhood of vortices.

Since the positional fluctuations of neighboring vortices are a
small-scale feature, it is sufficient to take into account disorder
within the Gaussian approximation (i.e. neglecting the dependence of
the pinning force on the vortex displacements), the large scale
properties of which have been examined in Eq. (\ref{W.Gauss}) above.
The disorder contribution to the relative displacement of two
neighbored vortices ${\bf r}_1$ and ${\bf r}_2$ separated by a bond
vector ${\bf a}={\bf r}_{12}$ is given by
\begin{eqnarray}
\Delta w({\bf a}) &=& \int_{\omega \bf q} 
\left(1 - e^{i \bf q \cdot a} \right)
\Delta C_{\alpha \alpha} ({\bf q},\omega)\nonumber \\
&\approx& \frac 12 \int_{\bf q} \sum_{\bf Q} 
\frac{({\bf a} \cdot {\bf q})^2 k^2 \Delta({\bf k})}
{\eta ^2 v^2 k_x^2 + c^2 q^4}
\end{eqnarray}
with $\Delta C$ taken from Eq. (\ref{C.Gauss}).

Now one can compare the shaking of a bond parallel to velocity (${\bf
a}=a \hat{\bf x}$) and ``perpendicular'' to the velocity (${\bf a}=a
\hat{\bf y}$). Strictly speaking, there are no bonds with ${\bf a}
\perp {\bf v}$ in a hexagonal lattice, by ``perpendicular'' we mean
the bonds making the 60$^o$ and/or 120$^o$ angles with the
velocity. This simplified treatment does not change our qualitative
conclusions altering slightly only the unimportant numerical
factors. In the limit of large velocities we find
\begin{mathletters}
\begin{eqnarray}
\Delta w(a \hat{\bf x}) &\approx & 
\frac {a \Delta_0}{\xi^{d+1} \eta^2 v^2} ~,\\
\Delta w(a \hat{\bf y}) &\approx & 
\frac {a^2 \Delta_0}{c \xi^{d+1} \eta v} ~.
\end{eqnarray}
\end{mathletters}
From this result one sees immediately that in this limit the bonds
perpendicular to ${\bf v}$ experience much stronger shaking effects
than the bonds parallel to velocity. Consequently, these perpendicular
bonds linking different channels are expected to break more easily
than the parallel bonds. This implies that the vortex structures has
much longer correlations in the direction parallel to the velocity
than in the other directions, in agreement with the anisotropy of the
dynamic Larkin lengths (\ref{larkin}). This result leads to the phase
diagram depicted in Fig. \ref{fig.phases} that has been discussed
already in the introduction.

Ultimately it is desirable to have a more systematic approach to the
effects of dislocations in the driven medium. A first step in this
direction, the study of the dynamics of single dislocations, was made
in Ref. \cite{SV97}. In order to decide whether {\em free}
dislocations destroying the topological order of the lattice are
present, it is necessary to study the dynamic stability of dislocation
loops (in $d=3$) or of dislocation pairs (in $d=2$). In particular in
$d=2$ one can expect a description of the dynamic phase transition in
terms of the Kosterlitz-Thouless transition \cite{KT73} generalized
to non-equilibrium systems.

\section{Conclusions}
\label{sec.conc}

We have constructed a general approach to the driven dynamics of dirty
periodic media based on the MSR technique. The developed scheme
provides a regular and consistent derivation of the effects of
disorder on the sliding motion. At present, however, we have
restricted ourselves to the second order dynamical perturbation
theory, yet sufficient to draw several fundamental conclusions
concerning the high-velocity behavior.

We have derived the coarse-grained equation of motion (\ref{eqmo^*})
for periodic media in the presence of disorder.  We have found a
renormalization of the bare system parameters like friction
coefficients, elastic constants, and the friction force within the
one-loop approximation as well as new couplings giving rise to the
disorder induced stresses, KPZ non-linearities and an effective
disorder with a random-force character evolving from the original
random potential \cite{BF95}. The presence of such terms has been
proposed for one-component systems like driven interfaces \cite{jK95}
and CDW \cite{BF95} without analytic derivation.

The appearance of divergent parameters under coarse-graining is much
more subtle in the driven system than in the system in equilibrium.
For complete coarse-graining ($\Lambda^< \to 0$) we found a divergence
in the correlator of the transverse components of the effective
thermal noise ${\bbox \xi}^*$ and the random pinning force ${\bf f}^*$
in $d \leq 3$. Thus in the driven state the upper critical dimension
is reduced by one and there are less divergent parameters in
comparison to the static case, where also the friction coefficients
and elastic constants diverge already in the first order of
perturbation theory for $d \leq 4$.

The divergences in the correlator of the (persistent and
non-persistent) random force components perpendicular to the drift
velocity appear only for systems with a periodic structure transverse
to the velocity. Therefore there is a fundamental difference between
the dynamic behavior of CDW, which have only one component, and VL
with more than one displacement components.

The standard way to test {\em glassy properties} of the systems in
question is to examine the large scale behavior of the
disorder-induced corrections to the physical quantities, the most
marking of which is the friction coefficient. Its divergence is
immediately related to an extremely slow dynamics that is dominated by
infinitely high barriers. A large-scale divergence of the perturbative
corrections would then indicate the glassy behavior.  However, a
divergence of the first order correction to the friction coefficient
is {\em absent} in the driven case.

Since the vortex system is already in motion, the friction coefficient
$\eta^*_{xx}$, which describes the velocity response for a change of
the {\em amplitude} of the driving force, has to be finite since the
drifting lattice already overcomes the potential barriers in this
direction. Nevertheless, one could expect that the friction
coefficient $\eta^*_{yy}$, which describes the velocity response for a
change of the {\em direction} of the driving force, could still
diverge due to infinite barriers for a transverse motion of the
lattice. However, since $\eta_{yy}^*$ is finite, these barriers can
only be finite. This implies a linear transverse transport
characteristic for small transverse forces at finite temperatures. It
is still possible that a true critical force exists at $T=0$. But this
is not a signature of glassy behavior. An instructive comparison is
provided by a single particle in a sinusoidal potential, which has no
glassy properties \cite{AH69}. Its transport characteristics has a
finite critical force at $T=0$. For low temperatures it has an
exponentially large but finite friction coefficient at small
velocities (due to thermal activation) that crosses over to the
smaller bare friction coefficient at large velocities.

Despite of the absence of generic glassy features like the existence of
transverse barriers, we find a tendency to form channels directly from
the presence of the stress coefficients (\ref{chi.xx}) in the
coarse-grained equation of motion.

To discuss the physical meaning of the effective temperature
$\vartheta^*$, it seems appropriate to emphasize that it is actually
the {\em weight of the noise correlator}. In a system at equilibrium
(and for our bare system), $\vartheta$ is proportional to the product
of the temperature $T$ and the friction coefficient $\eta$. Since we
have found a finite correction to the friction coefficient, the
divergence of $\vartheta$ can be interpreted as a divergence of the
effective temperature. However, in the non-equilibrium situation under
consideration, there is no well-defined meaning of a ``temperature.''
One can speak only about an (non-unique) effective temperature if one
specifies what physical property of the non-equilibrium system is
compared to an equilibrium system. Since the divergence of
$\vartheta^*$ arises from fluctuations on the largest length scales,
only the degrees of freedom on asymptotically large scale can be
related to an infinite effective temperature. This means that the
driven lattice in the presence disorder is on large length scales much
more rough than the lattice in the absence of disorder.

In general, divergent parameters signal a break-down of perturbation
theory at large scales. Therefore the question about the asymptotic
large-scale behavior of the system can be conclusively addressed only
by a systematic RG treatment that includes implicitly all orders of
perturbation theory, which goes beyond the scope of this work.

In the absence of a formal derivation, we propose the following
speculations regarding the existence of (quasi-) long-range order at
highest driving forces. It is essential to distinguish between
CDW-like systems (having only one ``displacement'' component) and
VL-like systems (having more than one ``displacement'' component)
because of fluctuations in the displacement components perpendicular
to the velocity. In the CDW case, where there is only the component
parallel to the velocity, it has been argued in Ref. \cite{BMR97} that
the random forces along that direction lead to a roughness with an
exponent $\zeta=(3-d)/2$ that is not reduced by renormalization
effects on the largest scales.

We believe that the situation could be different in the case of VL. We
have shown within the perturbative framework that the perpendicular
displacement components fluctuate much stronger than those of the
parallel component and are subject to a diverging random force
correlator. In this case one has to take into account that the strong
perpendicular displacements wash out not only the perpendicular
components of the pinning forces but also the parallel components and
therefore qualitatively {\em reduce} the true large-scale roughness in
all directions.

This speculation is formally supported by the structure of the
perturbative corrections obtained in Section \ref{sec.gen}. To be
specific, we discuss the random force correlator Eq. (\ref{phi}). It
contains exponential factors $\exp\{-{\bf k} \cdot {\bf W} \cdot {\bf
k} \}$ that have been neglected in the evaluation in Section
\ref{sec.eval}. This is legitimate in lowest order perturbation
theory, where only thermal fluctuations contribute to ${\bf
W}$. However, under iterating the perturbative expressions (this is
essentially the idea of an RG), one should take into account also the
disorder contributions to ${\bf W}$. Since we have found that disorder
roughens the system, ${\bf W}({\bf r},t)$ diverges on large scales and
the exponential factors decay for large ${\bf k}$, ${\bf r}$ or $t$,
suppressing the corrections in comparison to the lowest order
perturbative results. In this way all corrections generated by
disorder are suppressed, except from the first order correction
(\ref{Phi^1}) that tends to compensate the disorder itself.

One also immediately recognizes that the roughness of the components
perpendicular to the velocity (the large scale behavior of $W_{yy}$)
influence e.g. the random force correlator $\phi_{xx}$ in directions
parallel to the velocity and vice versa.

This mechanism is the same in all perturbative expressions and, as we
believe, coud reduce the true roughness qualitatively in comparison to
the perturbative result in $2 \leq d \leq 3$ for the {\em largest}
velocities.

In addition to the aspects of the high-velocity phase we have
evaluated our general perturbative results also in the limit of
vanishing velocities, when the depinning transition is approached. We
found disorder to be relevant in $d \leq 4$ in agreement with Larkin's
original static analysis \cite{aiL70} and dynamic approaches to
depinning as a critical phenomenon \cite{fisher,depin}. As an
additional feature our perturbative analysis revealed the relevance of
the non-equilibrium contributions ({\em diverging} KPZ terms and
random-force correlator) to the equation of motion.

To be specific, let us consider the example of the KPZ couplings as
true non-equilibrium couplings. It seems surprising that they do not
always vanish in the limit $v \to 0$, where one naively expects the
FDT to hold and all non-equilibrium terms to vanish. To resolve this
paradox note that the zero-velocity limit has to be taken with care
since it does not commute with the limit $\Lambda^< \to 0$. The
observed divergence occurs only if one takes $\Lambda^< \to 0$ before
$v \to 0$, since the divergence arises from the infrared contributions
to the integration over $q$. Physically these contributions are
related to the diverging energy barriers on large scales. These
diverging barriers imply a diverging relaxation time and persistent
memory effects of the system, which are the origin of the survival of
non-equilibrium terms.  In other words, in a glassy system relaxing
from a (globally drifting) non-equilibrium state into its equilibrium
state after switching off the current, there will be regions which are
still drifting, their dynamics being governed by an effective
non-equilibrium equation of motion.  If, on the other hand, one
considers the KPZ couplings for {\em finite} $\Lambda^<$, they are
finite and vanish $\sim v$ for velocities $v \ll c \xi {\Lambda^<}^2
/\eta$. For this reason these terms have not been taken into account
in the previous studies of the depinning transition. However, the
obervation that these non-equilibrium terms diverge at small
velocities for $\Lambda^<=0$ could indicate that even for weak
disorder the depinning transition is rounded in a very narrow region
by rare plastic effects \cite{Cop}, which are not captured by the
phenomenological Lindemann criterion.

In the final stage of preparing this manuscript we became aware of a
preprint by BMR \cite{BMR97long} that extends their earlier reference
\cite{BMR97} and where the authors come to similar conclusions. In
addition to perturbative results a RG analysis of a simplified ``toy
model'' for the transverse displacement component has been performed,
but it does not seem to yield results that differ qualitatively from
those obtained by the perturbation theory.

\section*{Acknowledgments}

We have benefited from stimulating discussions with I. Aronson, A.E.
Koshelev, T. Nattermann and L.H. Tang. We gratefully acknowledge a
critical reading of our manuscript by G. Crabtree.

This work was supported from Argonne National Laboratory through the
U.S.  Department of Energy, BES-Material Sciences, under contract No.
W-31-109-ENG-38 and by the NSF-Office of Science and Technology
Centers under contract No.  DMR91-20000 Science and Technology Center
for Superconductivity. S. acknowledges support by the Deutsche
Forschungsgemeinschaft project SFB341 and grant SCHE/513/2-1.

\appendix

\section{Notation}
\label{append.not}

For the convenience of the reader we summarize here our notation. The
total dimension is $d=d_\parallel + d_\perp$ with the inner vortex
dimension $d_\parallel$ and the number $d_\perp$ of displacement
components. Latin indices run over all $d$ components, whereas Greek
indices run over $d_\perp$ components. We use summation convention for
indices, i.e. ${\bf k} \cdot {\bf k} \equiv k_a k_a \equiv \sum_a
k_a^2$.
 
Some short notation:
\begin{mathletters}
  \begin{eqnarray}
    {\bf r}_{12} &:=& {\bf r}_1 - {\bf r}_2 ~, \quad
    t_{12} := t_1 - t_2
    \nonumber \\
    \delta_{12} &:=& \delta({\bf r}_{12}) \delta(t_{12}) ~, \quad 
    \delta({\bf r}) := a^{-d_\perp} \delta_{{\bf r}_\perp} 
    \delta({\bf r}_\parallel)
    \nonumber \\
    \int_i &:=& \int_{{\bf r}_i} \int_{t_i} ~,
    \quad \int_{t} := \int dt ~, 
    \quad \int_{{\bf r}} := a^d_\perp \sum_{{\bf r}_\perp}
    \int d^{d_\parallel} r_\parallel ~,
    \nonumber \\
    \int_{\bf k} &:=& \int \frac{ d^dk}{(2 \pi)^d} ~, \quad
    \int_{\bf q} := \int_{1BZ}\frac{ d^dq}{(2 \pi)^d} ~, \quad
    \int_\omega := \int\frac{ d\omega}{2 \pi} ~.
    \nonumber 
  \end{eqnarray}
\end{mathletters}
For discrete space-components the integration has to be replaced by a
sum with a factor $a^{d_\perp}$, the ``volume'' per vortex.  In
$d_\perp=2$ one has $a^2={\Phi_0}/{B}$.  ${\bf q}$-integrals run only
over the first Brillouin zone (1BZ) but ${\bf k}$ runs over whole
momentum space.

For two-point quantities a conjugation is defined by
\begin{eqnarray}
  {G^\dagger}_{\alpha \beta}({\bf k},\omega)&:=& 
  G_{\beta \alpha}(-{\bf k},-\omega) ~,
  \\
  {{\bf G}^\dagger}_{ij} &\equiv& 
  {G^\dagger}_{\alpha \beta}({\bf r}_i-{\bf r}_j,t_i-t_j):=
  G_{\beta \alpha}({\bf r}_j-{\bf r}_i,t_j-t_i) ~. \nonumber
\end{eqnarray}

\section{Second order correction to the disorder correlator}
\label{append.seco}

In order to find the second-order contributions to the disorder
correlator, one has to look at terms of order ${\cal O}({\tilde u}^2)$ in
(\ref{A.cumul}c), which we separate into two contributions (all
superscripts ``$\gtrless$'' are dropped again)
\begin{equation}
  \label{A^2_2}
  {\cal A}^{(2)}_{{\tilde u}^2}=
  {\cal A}^{(2,1)}_{{\tilde u}^2}+{\cal A}^{(2,2)}_{{\tilde u}^2}
\end{equation}
specified below. A further condensation of the products
\begin{equation}
  {\bbox \Phi}_{12}^{\bf k}:={\bbox \Phi}({\bf k}) 
  \ e^{i{\bf k} \cdot {\bf R}_{12} }
\end{equation}
and ${\Phi^\dagger}_{12\alpha \beta}^{\bf k}:=\Phi_{21\beta
  \alpha}^{-\bf k}$ helps us to keep the structure of the expressions
transparent: 
\beginwide
\begin{mathletters}
  \label{A^2_2.contr}
  \begin{eqnarray}
    {\cal A}^{(2,1)}_{{\tilde u}^2} &=&
    - \frac 12  \int_{1234 {\bf k}' {\bf k}''}
    \bigg\{
    [{\tilde {\bf u}}_1 \cdot {\bbox \Phi}_{13}^{\bf k'} \cdot 
    ({{\bf G}^\dagger}_{31} \cdot {\bf k}' 
    +({{\bf G}^\dagger}_{32}  -{{\bf G}^\dagger}_{34} ) \cdot {\bf k}'')]
%    \times \nonumber \\ && \quad
    [({\bf k}' \cdot ({\bf G}_{14} - {\bf G}_{34})
    +{\bf k}'' \cdot {\bf G}_{24})
    \cdot {{\bbox \Phi}^\dagger}^{-{\bf k}''}_{42} \cdot {\tilde {\bf u}}_2] 
    \ e^{-A}
    \nonumber \\ && \quad
    -[{\tilde {\bf u}}_1 \cdot {\bbox \Phi}_{13}^{\bf k'} 
    \cdot {{\bf G}^\dagger}_{31} \cdot {\bf k}']
    [{\bf k}'' \cdot {\bf G}_{24} 
    \cdot {{\bbox \Phi}^\dagger}^{-{\bf k}''}_{42} \cdot {\tilde {\bf u}}_2]
    \ e^{-{\bf k}' \cdot {\bf W}_{13} \cdot {\bf k}'
      -{\bf k}'' \cdot {\bf W}_{24} \cdot {\bf k}''}
    \bigg\} ~,
    \\
    {\cal A}^{(2,2)}_{{\tilde u}^2} &=&
    - \frac 14  \int_{1234 {\bf k}' {\bf k}''}
    [{\tilde {\bf u}}_1 \cdot {\bbox \Phi}^{\bf k'}_{12} 
    \cdot {\tilde {\bf u}}_2]
    ({\bf k}' \cdot ({\bf G}_{13} - {\bf G}_{23})
    -{\bf k}'' \cdot {\bf G}_{43})
    \cdot {\bbox \Phi}^{\bf k''}_{34} \cdot
%    \nonumber \\ && \quad
    (({{\bf G}^\dagger}_{41} -{{\bf G}^\dagger}_{42}) \cdot {\bf k}'
    +{{\bf G}^\dagger}_{43} \cdot {\bf k}'') 
    \ e^{-B} 
  \end{eqnarray}
\end{mathletters}
with the exponentials (generalize Debye-Waller factors)
\begin{mathletters}
  \begin{eqnarray}
    A &\equiv & A^{\bf k' k''}_{1234} := \frac 12
    \langle[{\bf k}' \cdot ({\bf u}_1 -{\bf u}_3)
    +  {\bf k}'' \cdot ({\bf u}_2 -{\bf u}_4)]^2 \rangle
    \nonumber \\
    &=& 
    {\bf k}' \cdot {\bf W}_{13} \cdot {\bf k}'
    +{\bf k}'' \cdot {\bf W}_{24} \cdot {\bf k}''
    -{\bf k}' \cdot {\bf W}_{12} \cdot {\bf k}''
    +{\bf k}' \cdot {\bf W}_{14} \cdot {\bf k}''
    +{\bf k}' \cdot {\bf W}_{32} \cdot {\bf k}''
    -{\bf k}' \cdot {\bf W}_{34} \cdot {\bf k}'' ~,
    \\
    B&\equiv& B^{\bf k' k''}_{1234} :=  A^{\bf k' k''}_{1324}  ~.
  \end{eqnarray}
\end{mathletters}
The second term in the curly brackets of Eq.  (\ref{A^2_2.contr}a)
arises from the subtraction in the definition of the cumulant $\langle
{\cal A}^{\rm pin}, {\cal A}^{\rm pin} \rangle_c:= \langle {\cal
  A}^{\rm pin} {\cal A}^{\rm pin} \rangle - \langle {\cal A}^{\rm pin}
\rangle ^2$, which correspond to ``disconnected diagrams'' in
field-theoretical language.  In Eq.  (\ref{A^2_2.contr}b) one further
term $\sim {{\bf G}^\dagger}_{34} {\bf G}_{34}$ actually vanishes due
to causality, i.e.  ${{\bf G}^\dagger}_{34} \sim \Theta(t_4>t_3)$ and
${\bf G}_{34} \sim \Theta(t_3>t_4)$.

Contributions that represent effective disorder are identified as
follows: in (\ref{A^2_2.contr}) the response fields are evaluated at
points $({\bf r}_i,t_i)$ with $i=1,2$. Disorder contributions are
those which persist for $|t_{12}| \to \infty$.  This requires that
${\tilde {\bf u}}_1$ and ${\tilde {\bf u}}_2$ are {\em not} connected
(even indirectly) by response functions. In this limit the factors
$e^{-A}$ and $e^{-B}$ simplify, since ${\bf W}_{ij} \to {\bf C}_0$ if
the points $i$ and $j$ are unconnected.  Therefore we are left with
\begin{mathletters}
\label{A^2p_2}
\begin{eqnarray}
  {\cal A}^{(2,1p)}_{{\tilde u}^2} &=&
  - \frac 12  \int_{1234 {\bf k}' {\bf k}''}
  {\tilde {\bf u}}_1 \cdot {\bbox \Phi}^{{\bf k}'}_{13} \cdot 
  \nonumber \\ && \quad \times
  \Big\{
  ({{\bf G}^\dagger}_{31} \cdot {\bf k}'  - {{\bf G}^\dagger}_{34} 
  \cdot {\bf k}'')
  ({\bf k}'  \cdot {\bf G}_{14} -  {\bf k}'  \cdot {\bf G}_{34})
  \ e^{-{\bf k}'' \cdot {\bf C}_0 \cdot {\bf k}''
    - {\bf k}' \cdot {\bf W}_{13} \cdot {\bf k}'
    - {\bf k}' \cdot ({\bf W}_{14}- {\bf W}_{34} )\cdot {\bf k}''}
  \nonumber \\ && \quad
  + ({{\bf G}^\dagger}_{32} \cdot {\bf k}''  - {{\bf G}^\dagger}_{34} 
  \cdot {\bf k}'')
  ({\bf k}''  \cdot {\bf G}_{24} -  {\bf k}'  \cdot {\bf G}_{34})
  \ e^{-{\bf k}' \cdot {\bf C}_0 \cdot {\bf k}'
    - {\bf k}'' \cdot {\bf W}_{24} \cdot {\bf k}''
    - {\bf k}'' \cdot ({\bf W}_{23}- {\bf W}_{43} )\cdot {\bf k}'}
  \nonumber \\ && \quad
  + {{\bf G}^\dagger}_{32} \cdot {\bf k}'' {\bf k}'  \cdot {\bf G}_{14}
  \ e^{-({\bf k}'-{\bf k}'') \cdot {\bf C}_0 \cdot ({\bf k}'-{\bf k}'')
    - {\bf k}' \cdot ({\bf W}_{14} +{\bf W}_{32})\cdot {\bf k}''}
  \Big\} \cdot {{\bbox \Phi}^\dagger}^{-{\bf k}''}_{42} 
  \cdot {\tilde {\bf u}}_2 ~.
\end{eqnarray}
Terms $\sim {{\bf G}^\dagger}_{31} {\bf G}_{24}$ are non-persistent
and therefore do no longer appear in ${\cal A}^{(2,1p)}_{{\tilde
    u}^2}$.  Vanishing terms $\sim {{\bf G}^\dagger}_{34} {\bf
  G}_{34}$ have been inserted by hand to complete the squares. The
second contribution becomes analogously
\begin{eqnarray}
  {\cal A}^{(2,2p)}_{{\tilde u}^2} &=&
  - \frac 14  \int_{1234 {\bf k}' {\bf k}''}
  [{\tilde {\bf u}}_1 \cdot {\bbox \Phi}^{\bf k'}_{12} 
  \cdot {\tilde {\bf u}}_2]
  \nonumber \\ && \quad
  \times \Big\{
  ({\bf k}'  \cdot {\bf G}_{13} - {\bf k}'' \cdot {\bf G}_{43})
  \cdot {\bbox \Phi}^{\bf k''}_{34} \cdot
  ({{\bf G}^\dagger}_{41} \cdot {\bf k}' + {{\bf G}^\dagger}_{43} 
  \cdot {\bf k}'')
  \ e^{-{\bf k}' \cdot {\bf C}_0 \cdot {\bf k}'
    -{\bf k}'' \cdot {\bf W}_{34} \cdot {\bf k}''
    +{\bf k}' \cdot ({\bf W}_{13} - {\bf W}_{14} ) \cdot {\bf k}''}
  \nonumber \\ && \quad
  +({\bf k}'  \cdot {\bf G}_{23} + {\bf k}'' \cdot {\bf G}_{43})
  \cdot {\bbox \Phi}^{\bf k''}_{34} \cdot
  ({{\bf G}^\dagger}_{42} \cdot {\bf k}' - {{\bf G}^\dagger}_{43} 
  \cdot {\bf k}'')
  \ e^{-{\bf k}' \cdot {\bf C}_0 \cdot {\bf k}'
    -{\bf k}'' \cdot {\bf W}_{34} \cdot {\bf k}''
    -{\bf k}' \cdot ({\bf W}_{23} - {\bf W}_{24} ) \cdot {\bf k}''}
  \nonumber \\ && \quad
  -  2 {\bf k}'  \cdot {\bf G}_{23} \cdot {\bbox \Phi}^{\bf k''}_{34} \cdot
  {{\bf G}^\dagger}_{41} \cdot {\bf k}' 
  \ e^{-({\bf k}' - {\bf k}'') \cdot {\bf C}_0 \cdot ({\bf k}' - {\bf k}'')
    -{\bf k}' \cdot ({\bf W}_{23} + {\bf W}_{14} ) \cdot {\bf k}''}
  \Big\} ~.
\end{eqnarray}
\end{mathletters}
\endwide
Here the number of terms was reduced using the relabeling
symmetry $3 \leftrightarrow 4$ with ${\bf k}'' \leftrightarrow -{\bf
  k}''$.

To identify the corrections to the disorder correlator we have to
analyze the dependence of the functional on the displacement field and
the response field. Since we are interested in the large-scale
physics, we can consider the response-functions as local in space and
time. The labels have been chosen such that ${\tilde {\bf u}}$ is
attached to 1 and 2. Since in all expressions two response-functions
are involved that connect point 3 and point 4 to point 1 or point 2,
${\bf u}_3$ and ${\bf u}_4$ may be replaced by the corresponding ${\bf
  u}_1$ or ${\bf u}_2$. These replacements can be considered as lowest
order of the expansion (\ref{appr.diff}).

To give an example we discuss the first term in the curly brackets of
Eq. (\ref{A^2p_2}a). There the response function connect points 3 and
4 with 1, whereas 2 is free. In this case it is convenient to replace
${\bf u}_3$ and ${\bf u}_4$ by ${\bf u}_1$. The exponential factors
implicit in the factors ${\bbox \Phi}^{\bf k}_{ij}$ then can be
rewritten as $e^{i{\bf k}' \cdot ({\bf r}_{13} + {\bf v} t_{13}) + i
  {\bf k}'' \cdot ({\bf r}_{24}+{\bf v} t_{24} + {\bf u}_{21})}=
e^{-i{\bf k}'' \cdot ({\bf r}_{12} + {\bf v} t_{12} + {\bf u}_{12}) +
  i {\bf k}' \cdot ({\bf r}_{13}+{\bf v} t_{13}) + i {\bf k}'' \cdot
  ({\bf r}_{14}+{\bf v} t_{14})}$.  Now the integrand has the same
functional dependence on the fields as the original disorder action
(\ref{A^i}) and a contribution to the correlator correction can be
identified. In the same way one can proceed with the other terms of
Eq.  (\ref{A^2p_2}) and finds (abbreviating ${\bf R}_{ij}^{(0)}:={\bf
  r}_{ij} + {\bf v} t_{ij}$) 
\beginwide
\begin{eqnarray}
  \label{Phi^2}
  &&{\bbox \Phi}^{(2)}({\bf k}) = - e^{-{\bf k} \cdot {\bf C}_0 \cdot {\bf k}}
  \ \int_{34 {\bf k}' {\bf k}''}
  \nonumber \\ && \quad \times
  \Big\{
  {\bbox \Phi}({{\bf k}'}) \cdot 
  \Big[
  ({{\bf G}^\dagger}_{31} \cdot {\bf k}'  - {{\bf G}^\dagger}_{34} 
  \cdot {\bf k}'')
  ({\bf k}'  \cdot {\bf G}_{14} -  {\bf k}'  \cdot {\bf G}_{34})
  \ e^{- {\bf k}' \cdot {\bf W}_{13} \cdot {\bf k}'
    - {\bf k}' \cdot ({\bf W}_{14}- {\bf W}_{34} ) \cdot {\bf k}''}
  \ e^{i {\bf k}' \cdot {\bf R}^{(0)}_{13}
    +i {\bf k}'' \cdot {\bf R}^{(0)}_{14}} 
  \delta({\bf k}+{\bf k}'')
  \nonumber \\ && \quad
  + ({{\bf G}^\dagger}_{32} \cdot {\bf k}''  - {{\bf G}^\dagger}_{34} 
  \cdot {\bf k}'')
  ({\bf k}''  \cdot {\bf G}_{24} -  {\bf k}'  \cdot {\bf G}_{34})
  \ e^{- {\bf k}'' \cdot {\bf W}_{24} \cdot {\bf k}''
    - {\bf k}'' \cdot ({\bf W}_{23}- {\bf W}_{43} )\cdot {\bf k}'}
  \ e^{i {\bf k}' \cdot {\bf R}^{(0)}_{23}
    +i {\bf k}'' \cdot {\bf R}^{(0)}_{24}} 
  \delta({\bf k}-{\bf k}')
  \nonumber \\ && \quad
  + {{\bf G}^\dagger}_{32} \cdot {\bf k}'' {\bf k}' \cdot {\bf G}_{14}
  \ e^{- {\bf k}' \cdot ({\bf W}_{14} +{\bf W}_{32}) \cdot {\bf k}''}
  \ e^{i {\bf k}' \cdot {\bf R}^{(0)}_{23}
    + i {\bf k}'' \cdot {\bf R}^{(0)}_{14}}
  \delta({\bf k}-{\bf k}'+{\bf k}'')
  \Big]  \cdot {{\bbox \Phi}^\dagger}(-{\bf k}'') 
  \nonumber \\ &&
  + \frac 12 {\bbox \Phi}({\bf k'}) 
  \Big[
  ({\bf k}'  \cdot {\bf G}_{13} - {\bf k}'' \cdot {\bf G}_{43})
  \cdot {\bbox \Phi}({\bf k''}) \cdot
  ({{\bf G}^\dagger}_{41} \cdot {\bf k}' + {{\bf G}^\dagger}_{43} 
  \cdot {\bf k}'')
  \ e^{-{\bf k}'' \cdot {\bf W}_{34} \cdot {\bf k}''
    +{\bf k}' \cdot ({\bf W}_{13} - {\bf W}_{14} ) \cdot {\bf k}''}
  \ e^{i{\bf k}'' \cdot {\bf R}^{(0)}_{34}}
  \delta({\bf k}-{\bf k}')
  \nonumber \\ && \quad
  + ({\bf k}'  \cdot {\bf G}_{23} + {\bf k}'' \cdot {\bf G}_{43})
  \cdot {\bbox \Phi}({\bf k''}) \cdot
  ({{\bf G}^\dagger}_{42} \cdot {\bf k}' - {{\bf G}^\dagger}_{43} 
  \cdot {\bf k}'')
  \ e^{-{\bf k}'' \cdot {\bf W}_{34} \cdot {\bf k}''
    -{\bf k}' \cdot ({\bf W}_{23} - {\bf W}_{24} ) \cdot {\bf k}''}
  \ e^{i{\bf k}'' \cdot {\bf R}^{(0)}_{34}}
  \delta({\bf k}-{\bf k}')
  \nonumber \\ && \quad
  - 2 {\bf k}'  \cdot {\bf G}_{23} \cdot {\bbox \Phi}({\bf k''}) \cdot
  {{\bf G}^\dagger}_{41} \cdot {\bf k}' 
  \ e^{-{\bf k}' \cdot ({\bf W}_{23} + {\bf W}_{14} ) \cdot {\bf k}''}
  \ e^{i{\bf k}'' \cdot ({\bf R}^{(0)}_{14}
    - {\bf R}^{(0)}_{23})} \delta({\bf k} - {\bf k}' + {\bf k}'')
  \Big] \Big\}
\end{eqnarray}
\endwide
which in fact does not depend on points 1 or 2, which can be
eliminated by substitutions for the points 3 and 4.

The symmetry ${\bbox \Phi}({\bf k}) = {{\bbox \Phi}^\dagger}({\bf k})$
given for the original disorder correlator is preserved after the
inclusion of the corrections.

In order to go beyond the locality approximation used above, one could
include the derivatives of Eq. (\ref{appr.diff}). This is not done
here, since a scaling analysis shows that the resulting terms will be
less relevant than the disorder correlator. However, along these lines
one can straightforwardly derive a random KPZ nonlinearity as
postulated by Krug \cite{jK95}.

\end{multicols}

\begin{references}
\bibitem{Lambda} This cutoff may be chosen in an anisotropic way, e.g.
  by introducing diffrent cutoffs parallel and perpendicular to the
  vortices, $q_\parallel \leq \Lambda_\parallel$ and $q_\perp \leq
  \Lambda_\perp$. This complication is avoided in the text, but its
  generalization is straightforward.
\bibitem{Ito} We use the so-called Ito calculus where the
  instantaneous response vanishes, see appendix C of \cite{BJW76}.
\bibitem{note.eta} The evaluation of Eq. (\ref{eta.x}) requires some
  care. The numerator of the integrand is a difference of two terms.
  Each term for itself would give rise to a divergence in $d \leq 3$.
  This divergence, which is of the same type as in the random force
  correlator, is canceled in the difference of the two terms.
\end{references}
\end{document}